# Nearest-neighbor resonating valence bonds in YbMgGaO$_4$

Yuesheng Li[1,2,*], Devashibhai Adroja[3,4], David Voneshen[3], Robert I. Bewley[3],

Qingming Zhang[2,5,6,†], Alexander A. Tsirlin[1], and Philipp Gegenwart[1]

[1] Experimental Physics VI, Center for Electronic Correlations and Magnetism, University of Augsburg, 86159 Augsburg, Germany

[2] Department of Physics, Renmin University of China, Beijing 100872, People's Republic of China

[3] ISIS Pulsed Neutron and Muon Source, STFC Rutherford Appleton Laboratory, Harwell Campus, Didcot, Oxfordshire, OX11 0QX, United Kingdom

[4] Highly Correlated Matter Research Group, Physics Department, University of Johannesburg, PO Box 524, Auckland Park 2006, South Africa

[5] Department of Physics and Astronomy, Shanghai Jiao Tong University, Shanghai 200240, People's Republic of China

[6] Collaborative Innovation Center of Advanced Microstructures, Nanjing 210093, People's Republic of China

Email: *yuesheng.man.li@gmail.com, †qmzhang@ruc.edu.cn

**Since its proposal by Anderson[1,2], resonating valence bonds (RVB) formed by a superposition of fluctuating singlet pairs have been a paradigmatic concept in understanding quantum spin liquids (QSL). Here, we show that excitations related to singlet breaking on nearest-neighbor bonds describe the high-energy part of the excitation spectrum in YbMgGaO$_4$, the effective spin-1/2 frustrated antiferromagnet on the triangular lattice[3-6], as originally considered by Anderson. By a thorough single-crystal inelastic neutron scattering (INS) study, we demonstrate that nearest-neighbor RVB excitations account for the bulk of the spectral weight above 0.5 meV. This renders YbMgGaO$_4$ the first experimental system where putative RVB correlations restricted to nearest neighbors are observed, and poses a fundamental question of how complex interactions on the triangular lattice conspire to form this unique many-body state.**

QSL is a long-sought exotic phase in condensed matter physics. It is intimately related to the problem of high-temperature superconductivity and may be instrumental in realizing topological quantum computation[1,2,7-10]. In a QSL, spins are highly entangled up to long-distances and times without symmetry breaking down to zero temperature due to strong quantum fluctuations[7]. Experimental systems exhibiting QSL behavior are actively sought after. However, most of the existing materials are suffering from magnetic defects[11,12], spatial coupling anisotropy[12-14], and (or) antisymmetric Dzyaloshinsky-Moriya anisotropy[15]. Recently, a triangular QSL candidate YbMgGaO$_4$ attracted much interest[3-6], because it seems to be free from all of the above effects. Neither spin freezing nor long-range ordering were detected by muon spin relaxation (μSR) down to 0.048 K[5]. Together with the absence of any residual spin entropy[3], this renders YbMgGaO$_4$ a unique material that may exhibit a gapless U(1) QSL ground state.

A QSL state can be represented by a superposition of many different partitions of a system into valence bonds (spin-0 singlet pairs)[7], as proposed by Anderson back in 1973[1,2]. Such valence bonds can be formed between nearest-neighbor spins and between spins beyond nearest neighbors. The longer the bond, the weaker the respective singlet pairing energy. Low-energy excitations arise from breaking long-range valence bonds or rearranging the short bonds into longer ones[7,16]. High-energy excitations result from breaking nearest-neighbor valence bonds. Therefore, for characterizing a QSL, the detailed investigation of both high- and low-energy excitations is required.

In YbMgGaO$_4$, excellent transparence with the optical gap exceeding ~ 3 eV and the robust insulating behavior with the unmeasurably high resistance suggest a large charge gap, placing the material deep in the Mott-insulator regime of the Hubbard model. Strong localization of the 4$f$ electrons of Yb$^{3+}$ should restrict magnetic interactions to nearest neighbors ($\mathbf{S}_1$ and $\mathbf{S}_2$), but these interactions are anisotropic[4],

$$H = J_{zz}S_1^z S_2^z + J_{\pm}(S_1^+ S_2^- + S_1^- S_2^+) + J_{\pm\pm}(\gamma_{12}S_1^+ S_2^+ + \gamma_{12}^* S_1^- S_2^-) \\ - \frac{iJ_{z\pm}}{2}(\gamma_{12}^* S_1^+ S_2^z - \gamma_{12} S_1^- S_2^z + < 1 \leftrightarrow 2 >), \quad (1)$$

owing to the strong spin-orbit coupling, where the local moment S = 1/2 is a pseudospin, i.e., a combination of spin and orbital moments[6,17-19]. The lowest-energy eigenstate of a dimer formed by such anisotropic pseudospins is, nevertheless, a pure singlet, $\frac{1}{\sqrt{2}}(|\uparrow\downarrow> - |\downarrow\uparrow>)$, with the energy -3/4J$_0$ for the antiferromagnetic isotropic coupling, $J_0 \equiv (4J_{\pm}+J_{zz})/3 = 0.13(1)$ meV[4], as observed experimentally. In contrast to Heisenberg spins, the Yb$^{3+}$ pseudospins do not form a three-fold degenerate triplet state and feature three non-degenerate excited states separated by 0.809J$_0$, 1.012J$_0$, and 1.179J$_0$ from the singlet state instead. Excitations of a system can be viewed as the transitions between the singlet ground state and one of the excited states. Therefore, the RVB picture holds, albeit with minor quantitative modifications due to the different structure of the excited states.

Two very recent INS studies reported a continuum of spin excitations in YbMgGaO$_4$ in the energy range between 0.25 and 1.5 meV[20,21], and a phenomenological interpretation of these excitations in terms of a spinon Fermi surface has been proposed[20]. However, given the nearest-neighbor magnetic energy of $J_0 = 0.13(1)$ meV only[4], the excitations were observed at energies between 2J$_0$ and 10J$_0$. Therefore, they are high-energy magnetic excitations of YbMgGaO$_4$.

Below, we propose a different interpretation of these high-energy excitations and also endeavor to probe YbMgGaO$_4$ at lower energies. This task is extremely challenging, owing to the low energy scale of J$_0$ and the limits of instrumental energy resolution for neutron spectrometers. We report a thorough INS investigation of a single crystal of YbMgGaO$_4$ at energies between 0.02 and 3.5 meV, i.e. 0.15 to 27 in units of J$_0$. We present the data collected at the low temperature of 0.1 K, which is well inside the gapless ground-state regime defined by the saturation of the μSR rate[5], and at a much higher temperature of 35 K corresponding to 23J$_0$. The high-energy excitations observed previously[20,21] are confirmed and ascribed to nearest-neighbor RVB correlations. At low temperatures, these excitations are suppressed at energies below J$_0$, which suggests their gapped nature. Our results imply that distinct gapless excitations should exist at much lower energies, and we indeed observe traces of such excitations at the lowest energies accessible in our experiment.

The INS data for YbMgGaO$_4$ are shown in Fig. 1 and Fig. 2. A continuum of excitations broadly distributed in both momentum (**Q**) (see Fig. 1) and energy (0.1 ≤ ℏω ≤ 2 meV) space (see Fig. 2) is clearly visible. At 0.1 K, external field shifts the spectral weight towards higher energies (see Fig. 2), thus indicating the magnetic origin of these excitations. Remarkably, the excitation continuum persists up to 35 K, i.e., at a temperature that is 23 times higher than $J_0$. In fact, there are no qualitative differences between the high-energy parts of the INS spectra measured at 0.1 and 35 K apart from a 2.57(4)-fold reduction in the intensity near the hump center ~ 0.7 meV (see Fig. 2) when the temperature is increased to 35 K. The wave-vector and temperature dependence of the excitation continuum clearly indicates its spin-spin correlation origin and excludes other possible interpretations, such as CEF excitations, which are **Q**-independent and observed at energies larger than 39 meV[4,6,21].

We first focus on the wave vector dependence of the INS intensity measured with the incident neutron energy of $E_i = 5.5$ meV. Assuming uncorrelated nearest-neighbor valence bonds on a triangular lattice, the equal-time INS intensity can be expressed as[22]

$$N|F(\mathbf{Q})|^2 = \frac{2}{3} N|f(\mathbf{Q})|^2 \{3 - \cos(2\pi H) - \cos(2\pi K) - \cos[2\pi(H+K)]\}. \quad (2)$$

Here, $f(\mathbf{Q})$ is the magnetic form factor of free $Yb^{3+}$, and N is the total number of nearest-neighbor valence bonds probed in the INS measurement. This expression accounts for the experimental spectral weight above 0.5 meV, thus suggesting that at high energies spin-spin correlations are restricted to nearest neighbors. Any static state, such as valence bond solid[23] and glass[24,25], is excluded by our previous µSR study[5], and the RVB scenario turns out to be most plausible, as supported by the following arguments:

I. The **Q**-dependence of the INS signal at 0.1 and 35 K (after the subtraction of the background term *b*) is well described by the uncorrelated nearest-neighbor valence bond model on a triangular lattice (see Fig. 1c-f). No signatures of spin-spin correlations beyond nearest neighbors are observed (see Supplementary Information). This **Q**-dependence cannot be understood by short distance correlations in an arbitrary ground state on the triangular lattice. For example, the 120 degree long-range order would produce spin-wave excitations[26] and a qualitatively different **Q**-dependence even at high energies (see Supplementary Information).

II. The antiferromagnetic nature of the isotropic nearest-neighbor coupling, $J_0 \equiv (4J_\pm + J_{zz})/3 = 0.13(1)$ meV[4], allows the formation of spin singlet in a pair of the $Yb^{3+}$ spins.

III. Temperature dependence of the pre-factor *a* in the RVB expression, $a(35K)/a(0.1K) \sim 0.3$ (see Supplementary Information), is consistent with the expected ratio,

$$\frac{N(T_2 = 35K)}{N(T_1 = 0.1K)} \sim \frac{1 + \exp\left(-\frac{0.809 J_0}{k_B T_1}\right) + \exp\left(-\frac{1.012 J_0}{k_B T_1}\right) + \exp\left(-\frac{1.179 J_0}{k_B T_1}\right)}{1 + \exp\left(-\frac{0.809 J_0}{k_B T_2}\right) + \exp\left(-\frac{1.012 J_0}{k_B T_2}\right) + \exp\left(-\frac{1.179 J_0}{k_B T_2}\right)} \sim 0.26, \quad (3)$$

based on the thermal distribution of the eigenstates of the $Yb^{3+}$ dimer. With increasing temperature, a larger fraction of nearest-neighbor singlets is excited.

IV. The uniform spin susceptibility, $\chi'(E)$, which is obtained from the INS spectrum measured around the Gamma point ($\mathbf{Q} = 0$) via the fluctuation-dissipation theorem and the Kramers-Kronig transformation[22], is almost zero at 0.1 K above ~ 0.5 meV, in agreement with the proposed RVB state (see Supplementary Information).

V. The energy dependence of the integrated INS signal reveals gapped nature of the high-energy excitations (see below for the details), which is consistent with the aforementioned suppression of the uniform susceptibility above ~ 0.5 meV (see IV).

VI. Both spin and valence bond freezing are excluded by our μSR measurement reported previously[5].

The arguments I to VI suggest that the whole excitation continuum at energies above $J_0$ may be due to the nearest-neighbor RVB-type correlations. We prove this explicitly above 0.5 meV, while below 0.5 meV the **Q**-dependent data measured with the incident energy $E_i = 5.5$ meV are contaminated by the elastic signal (see Supplementary Information). Lower energies can be probed with $E_i = 1.26$ meV (the energy resolution $\sigma \sim 20$ μeV[27]), but these data cover a limited **Q**-range only. Nevertheless, we find no qualitative differences between the spectra at ~ 0.3 and ~ 0.7 meV in all measured **Q** space ($E_i = 1.26$ meV) apart from an overall increase in the intensity. This indicates same, nearest-neighbor nature of spin-spin correlations across the whole excitation continuum above $J_0$ that was previously ascribed to the spinon Fermi surface.

It is crucial, though, that this continuum and the associated nearest-neighbor spin-spin correlations do not persist down to zero energy, because the nearest-neighbor RVBs are gapped, whereas YbMgGaO$_4$ clearly shows gapless behavior[3,5]. Therefore, the RVB scenario holds at high energies only. The presence of a distinct low-energy regime is supported by the analysis of the energy-dependent spectra integrated over all measured **Q** space.

For energy transfer below $J_0$, excitations related to the breaking of nearest-neighbor spin singlets must freeze out as long as thermal energy is insufficient to overcome $J_0$, i.e. T < 1.5 K. We, therefore, expect that below 0.13 meV the INS intensity at 0.1 K falls below that at 35 K. As indicated by the downward-pointing arrow in Fig. 3c, this expected crossing of the overall scattering intensity is observed indeed. Respectively, the intensity difference I(0.1 K)-I(35 K) at zero magnetic field changes sign and becomes negative at energy transfer below $J_0$ (Fig. 3d).

Further information is obtained from the INS spectra at finite magnetic fields applied along the crystallographic *c*-direction. At 8.5 T, which fully polarizes the moments at low temperatures[4,6], a clear boundary is observed in the low-energy magnetic excitations, leading to a crossing of I(0.1 K, 8.5 T) with I(35 K) near 1 meV, as indicated by the arrow in Fig. 2. This gap is related to the Zeeman energy[6,21] in the applied field of 8.5 T. In the same vein, under a moderate applied field of 1.8 T, which polarizes the spins only partially, negative values of I(0.1 K)-I(35 K) occur below 0.27 meV (see Fig. 3d). This energy lies in between $J_0$ and the Zeeman energy $\mu_0\mu_B g_\| H_\| = 0.39$ meV of spin-wave excitations for this field. When the field is reduced to zero, the crossing of intensities shifts to $J_0$ (Fig. 3c). We, therefore, associate this effect with an energy gap for the continuum of nearest-neighbor RVB-type excitations[7]. These excitations seem to be unrelated to the gapless spinon Fermi surface, in contrast to recent expectations based on the INS measurements at higher energies[20].

It is worth noting that a qualitatively similar crossing of the INS intensities measured at low and high temperatures has been recently observed in the frustrated pyrochlore Er$_2$Ti$_2$O$_7$[28], where magnetic excitations are gapped. In our case, the relation I(0.1 K, 0 T) < I(35 K, 0 T) is also clearly detected in zero field at transfer energies from 0.13 meV down to 0.018 meV, below which a rapid increase of the low-temperature intensity sets in. This lower energy is roughly the same as the energy resolution $\sigma \sim 20$ μeV $(0.15 J_0)$[27] of the LET spectrometer at the incident neutron energy of 1.26 meV. We emphasize that the inelastic signal does not become featureless at this energy ($\sigma$), as otherwise a smooth convoluted Lorentzian-Gaussian peak profile would be expected (see the raw data in Fig. 3c and Supplementary Information). At low transfer energies, the inelastic signal is found on top of the elastic background (see Fig. 3c). Assuming a weakly temperature dependent elastic signal at T ≤ 35 K, we expect that it cancels out when analyzing I(0.1 K)-I(35 K). Therefore, the intensity difference observed in zero field (see Fig. 3d) is intrinsic, as further confirmed by its tangible field dependence, and should reflect the onset of low-energy excitations related to longer-range correlations in YbMgGaO$_4$[3,5]. The most conspicuous effect of this change is the shift of the intensity maxima from the K-points in the high-energy regime to the M-points in the low-energy regime (see Supplementary Information), as also seen in the diffuse scattering reported by Paddison et al.[21]

The clear separation between the low[3,5]- and high-energy excitations in the spectrum of YbMgGaO$_4$ (see Fig. 3d) is interesting and unique, rendering YbMgGaO$_4$ distinct from QSL materials known to date, such as herbertsmithite[11,29], organic charge transfer salts[13,14], and Ca$_{10}$Cr$_7$O$_{28}$ reported recently[30]. The RVB scenario on the triangular-lattice was also discussed for the cluster magnet LiZn$_2$Mo$_3$O$_8$, where a spin-liquid state with both nearest-neighbor and next-nearest-neighbor correlations is formed[31-33]. It is also worth noting that the continuum of nearest-neighbor RVB excitations goes back to the original idea by Anderson[1] who argued that Heisenberg spins on the regular triangular lattice evade long-range magnetic order and form the nearest-neighbor RVB QSL state. Although Anderson's conjecture was not confirmed in later studies[34], the formation of a QSL on a triangular lattice with spatial anisotropy[35], next-nearest-neighbor couplings[36], and multiple-spin exchange[37] was identified in the recent literature. Whereas the multiple-spin exchange can be clearly excluded due to the strongly localized nature of the 4$f$ electrons of Yb$^{3+}$, two other effects are potentially relevant to YbMgGaO$_4$.

The presence of next-nearest-neighbor couplings is currently debated based on the modeling of the magnetic diffuse scattering[21,38]. Spatial anisotropy of nearest-neighbor couplings can be, at first glance, excluded, based on the three-fold symmetry of the crystal structure[3]. However, recent experiments[6,21], including our inelastic neutron scattering study[6] of crystal-field excitations of Yb$^{3+}$, pinpoint the importance of the Mg/Ga disorder that leads to variations in the local environment of Yb$^{3+}$. An immediate effect of this structural disorder is the distribution of $g$-values that manifests itself in the broadening of excitations in the fully polarized state, yet randomness of magnetic couplings resulting in local spatial anisotropy seems to be relevant too[6,18,21].

Our result suggest that the broad excitation continuum in YbMgGaO$_4$ reflects nearest-neighbor spin correlations and bears no obvious relation to spinons, a conclusion consistent with the absence of spinon or any other magnetic contribution to the thermal conductivity[39]. On the other hand, gapless nature of YbMgGaO$_4$ evidenced by the non-zero low-temperature susceptibility[3,5] and the power-law behavior of the magnetic specific heat[3] are indicative of a distinct low-energy regime that has been glimpsed in our experiment. These low-energy excitations are likely to contain crucial information on whether the ground state of YbMgGaO$_4$ is indeed a QSL, or a special case of the "disorder-induced mimicry of a spin liquid", as proposed recently[40].

**Methods**

Systematic neutron scattering experiments were carried out on a cold neutron multi-chopper spectrometer LET at the ISIS pulsed neutron and muon source. Incident energies of 26.8, 5.5, 2.3, and 1.26meV were chosen for both elastic and inelastic scattering with the energy resolution of 1400, 160, 48, and 20 µeV, respectively[27]. Large single crystals (~ 1 cm) of YbMgGaO$_4$ were grown by the floating zone technique reported previously[4]. The as grown rod

(~ 50 g) was cut into slices along the *ab*-plane (the easily cleavable direction). Ten best-quality *ab*-slices of the single-crystal (total mass ~ 10 g) were selected for the neutron scattering experiment on LET by Laue X-ray diffractions on all surface (see Supplementary Information). The slices were fixed to the copper base by CYTOP glue to avoid any shift in an applied magnetic field up to 8.5 T. The sample temperature of 0.1 K was achieved using dilution refrigerator. The neutron diffraction (elastic signal) showed that the alignment of the single crystals was sufficient for the inelastic neutron scattering study of the continuous excitations. No additional diffraction peaks were observed down to 0.1 K, compatible with the absence of long-range magnetic order (see Supplementary Information). All neutron scattering data were processed and analyzed using Horace-Matlab[41] on the ISIS computers. Asymmetry of the intensities was observed due to the macro-scale non-rotational symmetry of the sample around the rotation axis. For the sake of clarity, the raw data have been symmetrized and averaged using the point symmetry ($D_{3d}$) in the reciprocal lattice space (see Fig. 1a, b). The corresponding raw data can be found in the Supplementary Information.

External magnetic fields of 1.8 and 8.5 T were applied along the *c*-axis. The data sets in Fig. 1a, b were integrated over the momentum space, $-0.9 \leq \eta \leq 0.9$ in $[0, 0, -\eta]$, and over a small energy range, $0.65 \leq E \leq 0.75$ meV. The data sets in Fig. 1c-e were integrated over the momentum space, $-1.03 \leq \xi \leq -0.97$ in $[\xi, -\xi/2, 0]$, $-0.03 \leq \xi \leq 0.03$ in $[\xi/2, -\xi, 0]$, and $-0.03 \leq \xi \leq 0.03$ in $[0, \xi, 0]$, respectively. All data sets in Fig. 1c-e were integrated over the same momentum range, $-0.9 \leq \eta \leq 0.9$ in $[0, 0, -\eta]$, and over the same energy range, $0.5 \leq E \leq 1.5$ meV. *a* and *b* are fitted constants for the proportionality and background, respectively (see Fig. 1c-e, and Supplementary Information). The data sets in Fig. 2 and Fig. 3 were integrated over all measured momentum space.

### Acknowledgements

We thank Gang Chen, Haijun Liao, Changle Liu, Sasha Chernyshev, and Mike Zhitomirsky for helpful discussions. Y. L. was supported by the start-up funds of Renmin University of China. Q. Z. was supported by the Fundamental Research Funds for the Central Universities, and by the Research Funds of Renmin University of China. This work was supported by the NSF of China (No. 11474357) and the Ministry of Science and Technology of China (973 Project No. 2016YFA0300504). The work in Augsburg was supported by the German Science Foundation through TRR-80 and the German Federal Ministry for Education and Research through the Sofja Kovalevskaya Award of the Alexander von Humboldt Foundation.

### Author Contributions

Y. L., D. A. and Q. Z. planned the experiments. Y. L. synthesized and characterized the sample. Y. L., D. V., R. I. B. and Q. Z. collected the neutron scattering data. Y. L. analyzed the data. Y. L., A. A. T. and P. G. wrote the manuscript with comments from all co-authors. The manuscript reflects the contributions of all authors.

### Author Information

The authors declare no competing financial interests. Correspondence and request for materials should be addressed to Y. L. (yuesheng.man.li@gmail.com).

**References**

[1] Anderson, P. W. Resonating valence bonds: a new kind of insulator? *Mater. Res. Bull.* **8**, 153-160 (1973).

[2] Anderson, P. W. The resonating valence bond state in $La_2CuO_4$ and superconductivity. *Science* **235**, 1196-1198 (1987).

[3] Li, Y. S. *et al.* Gapless quantum spin liquid ground state in the two-dimensional spin-1/2 triangular antiferromagnet $YbMgGaO_4$. *Scientific Reports* **5**, 16419 (2015).

[4] Li, Y. S. *et al.* Rare-earth triangular lattice spin liquid: a single-crystal study of $YbMgGaO_4$. *Phys. Rev. Lett.* **115**, 167203 (2015).

[5] Li, Y. S. *et al.* Muon spin relaxation evidence for the U(1) quantum spin-liquid ground state in the triangular antiferromagnet $YbMgGaO_4$. *Phys. Rev. Lett.* **117**, 097201 (2016).

[6] Li, Y. S. *et al.* Crystalline electric-field randomness in the triangular lattice spin-liquid $YbMgGaO_4$. *Phys. Rev. Lett.* **118**, 107202 (2017).

[7] Balents, L. Spin liquids in frustrated magnets. *Nature* **464**, 199-208 (2010).

[8] Lee, P. A. An end to the drought of quantum spin liquids. *Science* **321**, 1306 (2008).

[9] Wen, X. G. Quantum field theory of many-body systems: from the origin of sound to an origin of light and electrons (Oxford University Press, New York, 2004).

[10] Nayak, C., Simon, S. H., Stern, A., Freedman, M., Das Sarma, S. Non-Abelian anyons and topological quantum computation. *Rev. Mod. Phys.* **80**, 1083-1159 (2008).

[11] Lee, S. H. *et al.* Quantum-spin-liquid states in the two-dimensional kagome antiferromagnets $Zn_xCu_{4-x}(OD)_6Cl_2$. *Nat. Mater.* **6**, 853 (2007).

[12] Li, Y. S. *et al.* Gapless quantum spin liquid in the s = 1/2 anisotropic kagome antiferromagnet $ZnCu_3(OH)_6SO_4$. *New J. Phys.* **16**, 093011 (2014).

[13] Shimizu, Y., Miyagawa, K., Kanoda, K., Maesato, M. & Saito, G. Spin liquid state in an organic Mott insulator with a triangular lattice. *Phys. Rev. Lett.* **91**, 107001 (2003).

[14] Itou, T., Oyamada, A., Maegawa, S., Tamura, M. & Kato, R. Quantum spin liquid in the spin-1/2 triangular antiferromagnet $EtMe_3Sb[Pd(dmit)_2]_2$. *Phys. Rev. B* **77**, 104413 (2008).

[15] Zorko, A. *et al.* Dzyaloshinsky-Moriya anisotropy in the spin-1/2 kagome compound $ZnCu_3(OH)_6Cl_2$. *Phys. Rev. Lett.* **101**, 026405 (2008).

[16] Kalmeyer, V. & Laughlin, R. B. Equivalence of the resonating-valence-bond and fractional quantum Hall states. *Phys. Rev. Lett.* **59**, 2095 (1987).

[17] Ross, K. A., Savary, L., Gaulin, B. D. & Balents, L. Quantum excitations in quantum spin ice. *Phys. Rev. X* **1**, 021002 (2011).

[18] Gaudet, J. *et al.* Neutron spectroscopic study of crystalline electric field excitations in stoichiometric and lightly stuffed $Yb_2Ti_2O_7$. *Phys. Rev. B* **92,** 134420 (2015).

[19] Li, Y. D., Wang, X. & Chen, G. Anisotropic spin model of strong spin-orbit-coupled triangular antiferromagnets, *Phys. Rev. B* **94**, 035107 (2016).

[20] Yao, S. *et al.* Evidence for a spinon Fermi surface in a triangular-lattice quantum-spin-liquid candidate. *Nature* **540**, 559-562 (2016).

[21] Paddison, J. A. M. *et al.* Continuous excitations of the triangular-lattice quantum spin liquid $YbMgGaO_4$. *Nat. Phys.* **13**, 117-122 (2017).

[22] Xu, G., Xu, Z. & Tranquada, J. M. Absolute cross-section normalization of magnetic neutron scattering data. *Rev. Sci. Instrum.* **84**, 083906 (2013).

[23] Matan, K. *et al.* Pinwheel valence-bond solid and triplet excitations in the two-dimensional deformed kagome lattice. *Nat. Phys.* **6**, 865-869 (2010).

[24] Tarzia, M. & Biroli, G. The valence bond glass phase. *Europhys. Lett.* **82**, 67008 (2008).

[25] Singh, R. R. P. Valence bond glass phase in dilute kagome antiferromagnets. *Phys. Rev. Lett.* **104**, 177203 (2010).

[26] Toth, S. & Lake, B. Linear spin wave theory for single-Q incommensurate magnetic structures. *J. Phys.: Condens. Matter* **27**, 166002 (2015).

[27] Bewley, R. I., Taylor, J. W. & Bennington, S. M. LET, a cold neutron multi-disk chopper spectrometer at ISIS. *Nuclear Instruments and Methods in Physics* **637** 128-134 (2011).

[28] Ross, K. A., Qiu, Y., Copley, J. R. D., Dabkowska, H. A. & Gaulin, B. D. Order by Disorder Spin Wave Gap in the XY Pyrochlore Magnet $Er_2Ti_2O_7$. *Phys. Rev. Lett.* **112**, 057201 (2014).

[29] Han, T. H. *et al.* Fractionalized excitations in the spin-liquid state of a kagome-lattice antiferromagnet. *Nature* **492**, 406-410 (2012).

[30] Balz, C. *et al.* Physical realization of a quantum spin liquid based on a complex frustration mechanism. *Nat. Phys.* **12**, 942-949 (2016).

[31] Sheckelton, J. P., Neilson, J. R., Soltan, D. G. & McQueen, T. M. Possible valence-bond condensation in the frustrated cluster magnet $LiZn_2Mo_3O_8$. *Nat. Mater.* **11**, 493-496 (2012).

[32] Sheckelton, J. P. *et al.* Local magnetism and spin correlations in the geometrically frustrated cluster magnet $LiZn_2Mo_3O_8$. *Phys. Rev. B* **89**, 064407 (2014).

[33] Mourigal, M. *et al.* Molecular quantum magnetism in $LiZn_2Mo_3O_8$. *Phys. Rev. Lett.* **112**, 027202 (2014).

[34] Capriotti, L., Trumper, A. E. & Sorella, S. Long-range Néel order in the triangular Heisenberg model. *Phys. Rev. Lett.* **82**, 3899-3902 (1999).

[35] Yunoki, S. & and Sorella, S. Two spin liquid phases in the spatially anisotropic triangular Heisenberg model. *Phys. Rev. B* **74**, 014408 (2006).

[36] Kaneko, R., Morita, S. & Imada, M. Gapless spin-liquid phase in an extended spin 1/2 triangular Heisenberg model. *J. Phys. Soc. Jpn.* **83,** 093707 (2014).

[37] Misguich, G., Lhuillier, C., Bernu, B. & Waldtmann, C. Spin-liquid phase of the multiple-spin exchange Hamiltonian on the triangular lattice. *Phys. Rev. B* **60,** 1064–1074 (1999).

[38] Li, Y., Shen, Y., Li, Y., Zhao, J. & Chen, G. The effect of spin-orbit coupling on the effective-spin correlation in YbMgGaO$_4$. Preprint at https://arxiv.org/abs/1608.06445.

[39] Xu, Y. *et al.* Absence of magnetic thermal conductivity in the quantum spin-liquid candidate YbMgGaO$_4$. *Phys. Rev. Lett.* 117, 267202 (2016).

[40] Zhu, Z., Maksimov, P. A., White, S. R. & Chernyshev, A. L. Disorder-induced mimicry of a spin liquid in YbMgGaO$_4$. Preprint at https://arxiv.org/abs/1703.02971.

[41] Ewings, R. A., Buts, A., Le, M. D., van Duijn, J., Bustinduy, I. & Perring, T. G. HORACE: software for the analysis of data from single crystal spectroscopy experiments at time-of-flight neutron instruments. *Nuclear Instruments and Methods in Physics Research Section A* **834**, 132–142 (2016).

**Figure legends**

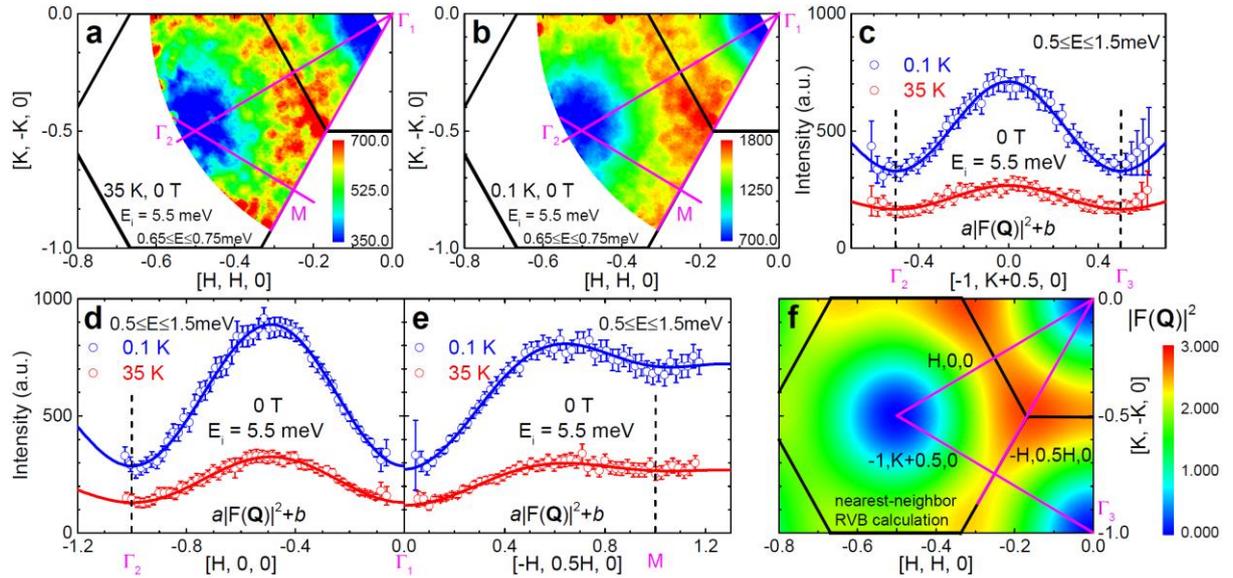

**Figure 1. Wave-vector dependences of the INS intensity for YbMgGaO$_4$ with comparison to the nearest-neighbor RVB model.** Wave-vector dependences of excitations measured under 0 T at 35 K (**a**) and at 0.1 K (**b**). Wave-vector dependences of the INS intensity along [-1, K+0.5, 0] (**c**), [H, 0, 0] (**d**), and [-H, 0.5H, 0] (**e**), with lines representing the calculated nearest-neighbor RVB dependence. Calculated $|F(\mathbf{Q})|^2$ (from Eq. (2)) (**f**). The black lines represent Brillouin zone boundaries. Pink lines show the high-symmetry directions with special reciprocal-space points labeled. Note that the experimental data contain a **Q**-independent background, which is about the same at $\Gamma_1$, $\Gamma_2$, and $\Gamma_3$. This background is missing in the RVB calculation in panel f, where $I_\Gamma = 0$.

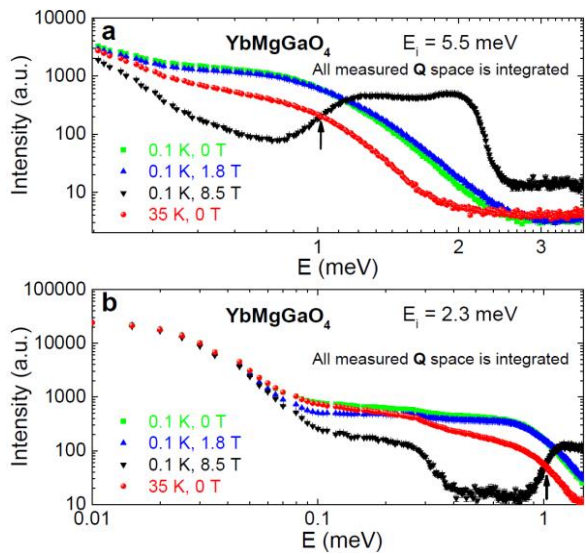

**Figure 2. Energy dependence of the integrated INS intensity for YbMgGaO$_4$.** The excitation continuum was probed with the incident neutron energy of 5.5 meV (**a**) and 2.3 meV (**b**). Black arrows show the lower boundary (gap energy) of the spin-wave excitations in the fully polarized state. This boundary is determined as the energy, where the high-temperature (35 K, 0 T) INS intensity crosses the low-temperature (0.1 K, 8.5 T) one.

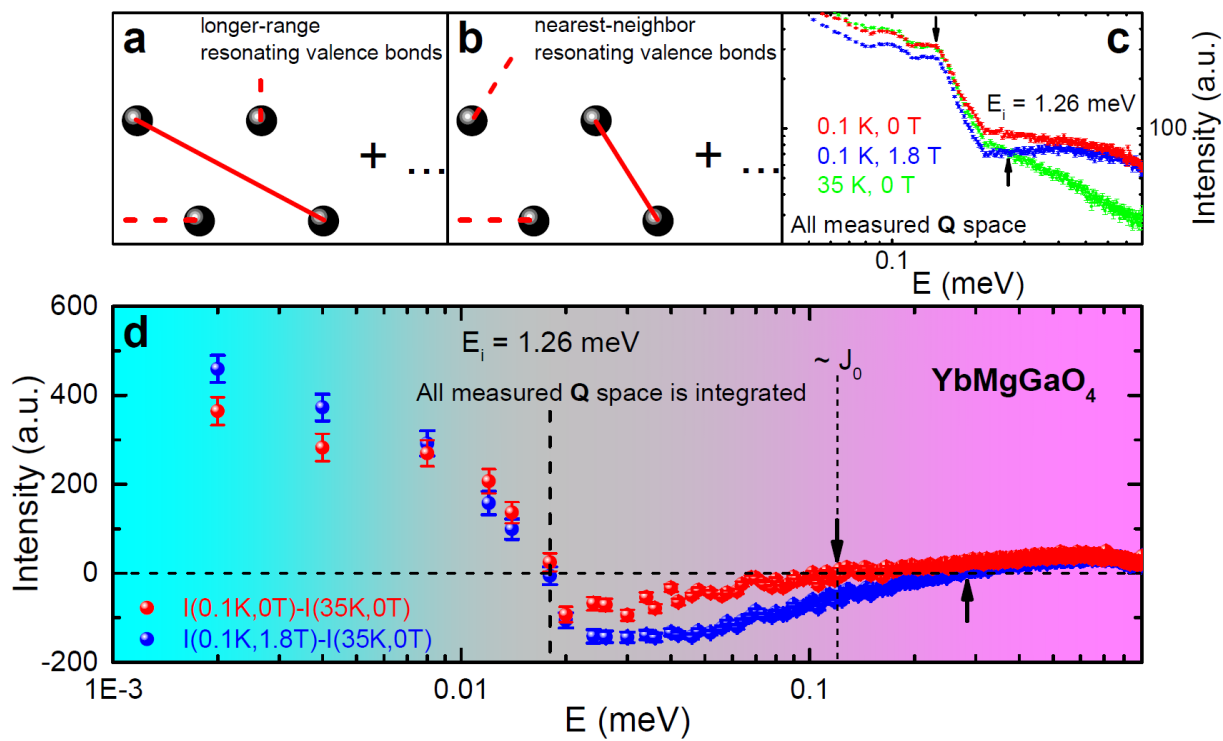

**Figure 3. INS data for YbMgGaO$_4$ with the lowest incident neutron energy of 1.26 meV. a.** and **b.** sketch of longer-range and nearest-neighbor RVBs on the triangular lattice (for one unit cell). The red lines represent the valence bonds (spin-singlets). **c**: Raw INS spectra **d**: integrated INS intensities at 0.1 K under 0 T and 1.8 T, after the subtraction by the corresponding spectrum measured at 35 K (0 T). Black arrows show the lower boundaries of the continuum excitations. The dashed vertical line indicates the low-energy crossing of the intensities.

# Supplementary Information for

## Nearest-neighbor resonating valence bonds in YbMgGaO$_4$

We present here:

**Fig. S1** Triangular lattice of Yb$^{3+}$ in YbMgGaO$_4$, and the diagonalization calculation for the Yb$^{3+}$ dimer with the anisotropic coupling.

**Fig. S2** Laue X-ray diffraction pattern.

**Fig. S3** Single-crystal sample for the neutron scattering measurements.

**Fig. S4** Elastic neutron diffraction pattern at 35 K and 0 T ($E_i$ = 26.8 meV).

**Fig. S5** Elastic neutron diffraction pattern at 0.1 K and 0 T ($E_i$ = 26.8 meV).

**Fig. S6** Elastic neutron diffraction pattern at 0.1 K and 1.8 T ($E_i$ = 26.8 meV).

**Fig. S7** Raw inelastic neutron scattering (INS) data at 35 K and 0 T ($E_i$ = 5.5 meV).

**Fig. S8** Raw INS data at 0.1 K and 0 T ($E_i$ = 5.5 meV).

**Fig. S9** Raw INS data at 0.1 K and 1.8 T ($E_i$ = 5.5 meV).

**Fig. S10** Calculated equal-time scattering intensities by uncorrelated second- and third-nearest-neighbor valence bonds on the triangular lattice. Both calculations are clearly very different from our experimental observations.

**Fig. S11** Wave-vector dependences of the INS intensity compared to different uncorrelated valence bond models, $a|F(\mathbf{Q})|^2+b$.

**Tab. S1** The fitted pre-factors $a$ and background constants $b$ using the nearest-neighbor RVB model.

**Fig. S12** Energy dependences of the uniform susceptibilities measured by INS at the Gamma point.

**Fig. S13** Temperature dependence of the static bulk susceptibilities and its comparison to the uniform susceptibilities.

**Fig. S14** Energy dependence of the integrated neutron scattering intensity and the elastic component ($E_i$ = 5.5 meV).

**Fig. S15** Energy dependence of the integrated neutron scattering intensity and the elastic component ($E_i$ = 1.26 meV).

**Fig. S16** Wave-vector dependence of the INS intensity integrated over $0.25 \leq E \leq 0.35$ meV, with $E_i$ = 5.5 meV.

**Fig. S17** Wave-vector dependence of the INS intensity integrated over $0.07 \leq E \leq 0.11$ meV, $0.28 \leq E \leq 0.32$ meV, and $0.68 \leq E \leq 0.72$ meV respectively with $E_i$ = 1.26 meV.

**Fig. S18** Wave-vector dependence of the elastic scattering signal ($-0.1 \leq E \leq 0.1$ meV) with $E_i$ = 5.5 and 1.26 meV.

**Fig. S19** Wave-vector dependences of the subtracted INS intensity [I(0.1K) – I(35K)] integrated over $0.07 \leq E \leq 0.11$ meV with $E_i$ = 1.26 meV.

**Fig. S20** Raw energy-dependent INS spectra with the incident neutron energy of 1.26 meV.

**Fig. S21-S23** Spin-wave excitations for the hypothetical 120 degree magnetic structure based on the experimental parametrization of the YbMgGaO$_4$ Hamiltonian, compared to the experimental data and to the model of nearest-neighbor RVB correlations.

**Fig. S24-S26** Spin-wave excitations for the 120 degree magnetically ordered ground state based on the ideal triangular-lattice spin-1/2 antiferromagnetic (J > 0) Heisenberg Hamiltonian, compared to the experimental data and to the model of nearest-neighbor RVB correlations.

The crystal field ground state of Yb$^{3+}$ in YbMgGaO$_4$ has been studied in our recent high-energy INS work [arxiv: 1702.01981 (accepted in Phys. Rev. Lett.)]. The wavefunctions of the ground-state Kramers doublet are ~ $\pm 0.71|\pm 7/2\rangle \mp 0.36|\mp 5/2\rangle + 0.60|\pm 1/2\rangle$. The effective spin-1/2 g-factors ($g_\parallel = 3.72$ and $g_\perp = 3.06$) had also been reported in our previously reported work [Phys. Rev. Lett. 115, 167203 (2015)].

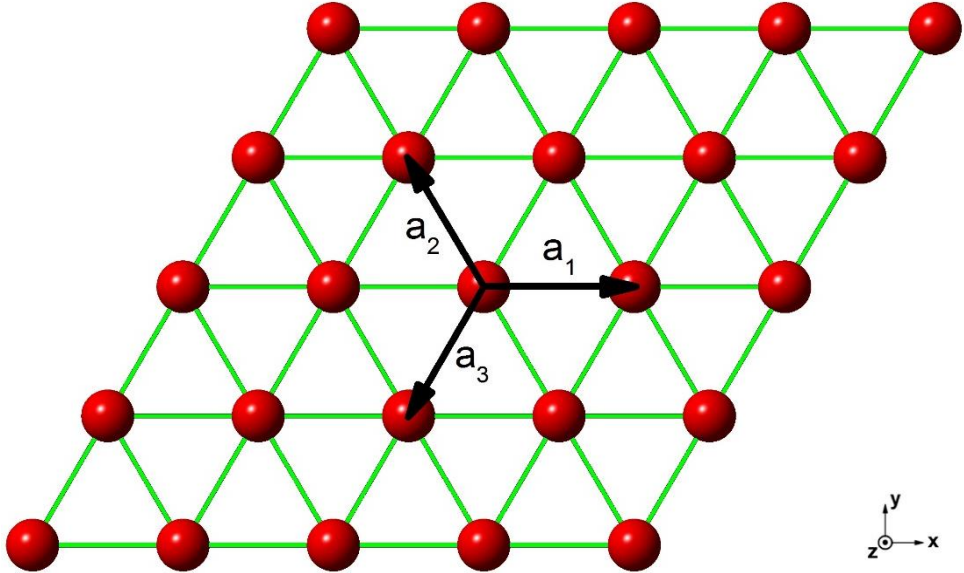

**Figure S1.** Triangular lattice of Yb$^{3+}$ in YbMgGaO$_4$.

The effective spin-1/2 Hamiltonian for the Yb$^{3+}$ dimer ($\mathbf{S}_1$ and $\mathbf{S}_2$) had been reported in our previous work [Phys. Rev. Lett. 115, 167203 (2015)].

$$H = J_{zz}S_1^z S_2^z + J_\pm(S_1^+ S_2^- + S_1^- S_2^+) + J_{\pm\pm}(\gamma_{12} S_1^+ S_2^+ + \gamma_{12}^* S_1^- S_2^-)$$
$$- \frac{iJ_{z\pm}}{2}(\gamma_{12}^* S_1^+ S_2^z - \gamma_{12} S_1^- S_2^z + <1 \leftrightarrow 2>). \qquad \text{Eq. S1}$$

Here, $J_\pm \sim 0.9$ K, $J_{zz} \sim 0.98$ K, $J_{z\pm} \sim 0$ K, $J_{\pm\pm} \sim$ (+ or -)0.155 K, and $\gamma_{12} = 1$, $e^{i2\pi/3}$, $e^{-i2\pi/3}$ for the bonds along the $\mathbf{a}_1$, $\mathbf{a}_2$, $\mathbf{a}_3$ directions (see Fig. S1), respectively.

Through the diagonalization of the Hamiltonian (Eq. S1), we find that the ground state is the *strict* traditional singlet, $\frac{1}{\sqrt{2}}(|\uparrow\downarrow\rangle - |\downarrow\uparrow\rangle)$. It is independent on the bond direction ($a_1$, $a_2$, or $a_3$) and the anisotropic coupling parameters, as long as the isotropic coupling, $J_0 \equiv (4J_\pm + J_{zz})/3$, is antiferromagnetic (i.e. $J_0 > 0$). And $J_0 = 1.5(1)$ K or $0.13(1)$ meV $> 0$ had been safely determined in our previous work [Phys. Rev. Lett. 115, 167203 (2015)].

The energy separations between the ground-state singlet (with the eigenvalue of energy, $-3/4J_0$) and three excited states are also independent of the bond direction ($a_1$, $a_2$, or $a_3$), but are strongly dependent on the anisotropic coupling parameters. Since $J_{z\pm}$ had been measured to be almost zero [Phys. Rev. Lett. 115, 167203 (2015)], all of the three excited energies (from $-3/4J_0$) are independent on the sign of $J_{\pm\pm}$ at $J_{z\pm} = 0$, and are calculated to be $0.809J_0$, $1.012J_0$, and $1.179J_0$, respectively.

As a result, the uncorrelated spin-1/2 nearest-neighbor valence bond model, which had been used for herbertsmithite [Nature 492, 406-410 (2012)], is applicable to YbMgGaO$_4$ as well.

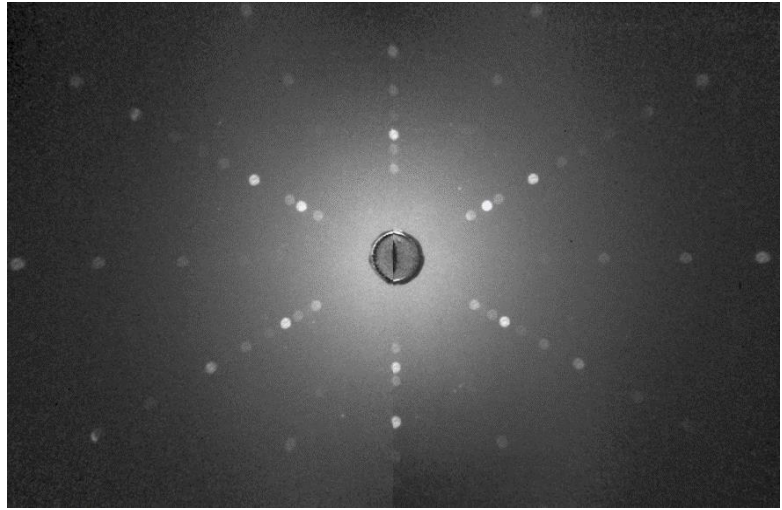

**Figure S2.** Laue X-ray diffraction pattern for the (0, 0, 1) surface of the YbMgGaO$_4$ single crystal.

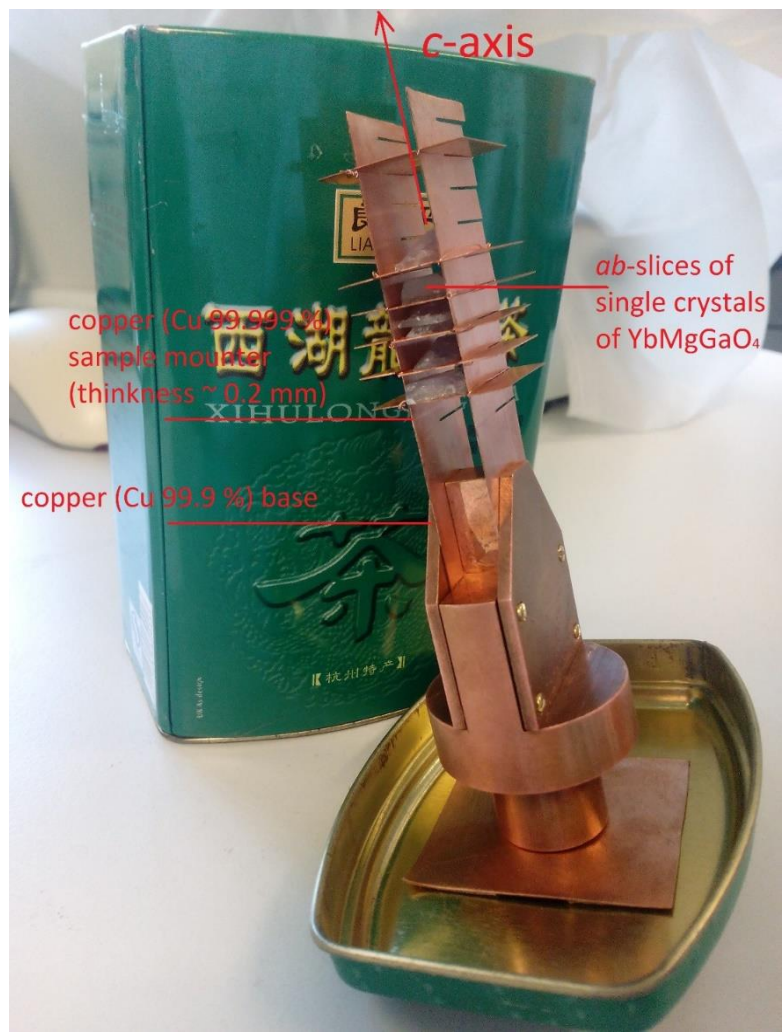

**Figure S3.** YbMgGaO$_4$ single-crystal sample (including a self-made sample mounter) used in the neutron scattering measurements.

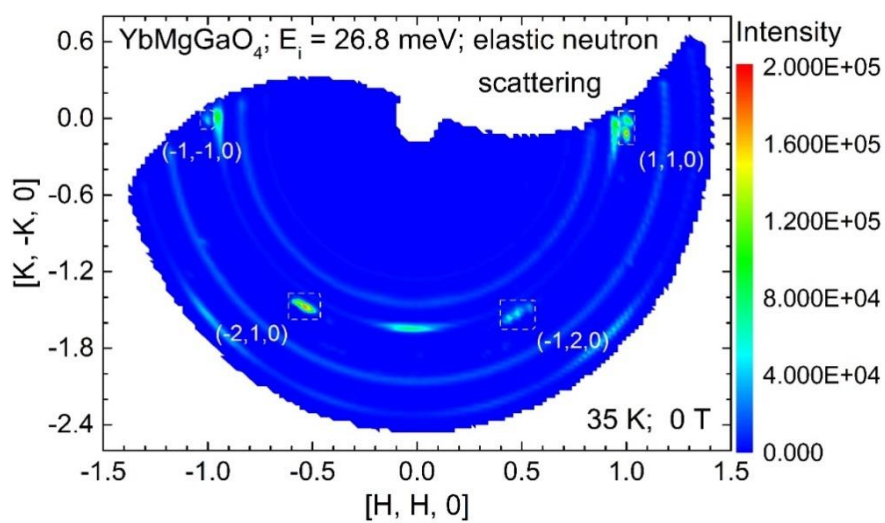

**Figure S4.** Neutron diffraction pattern for the YbMgGaO$_4$ single-crystal sample at 35 K and 0 T ($E_i$ = 26.8 meV).

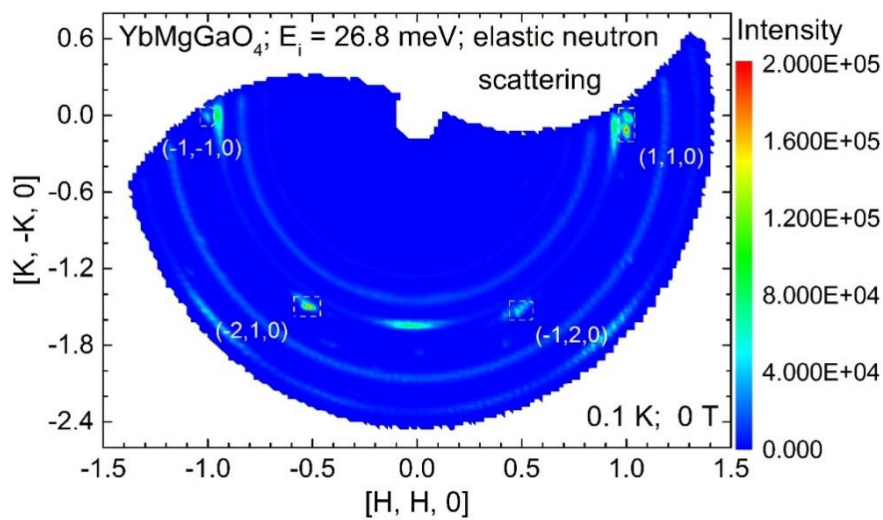

**Figure S5.** Neutron diffraction pattern for the YbMgGaO$_4$ single-crystal sample at 0.1 K and 0 T ($E_i$ = 26.8 meV).

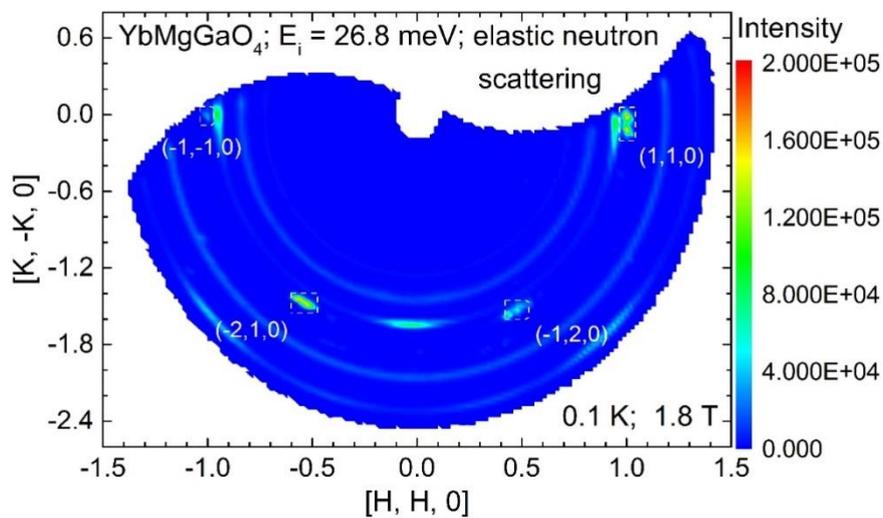

**Figure S6.** Neutron diffraction pattern for the YbMgGaO$_4$ single-crystal sample at 0.1 K and 1.8 T ($E_i$ = 26.8 meV).

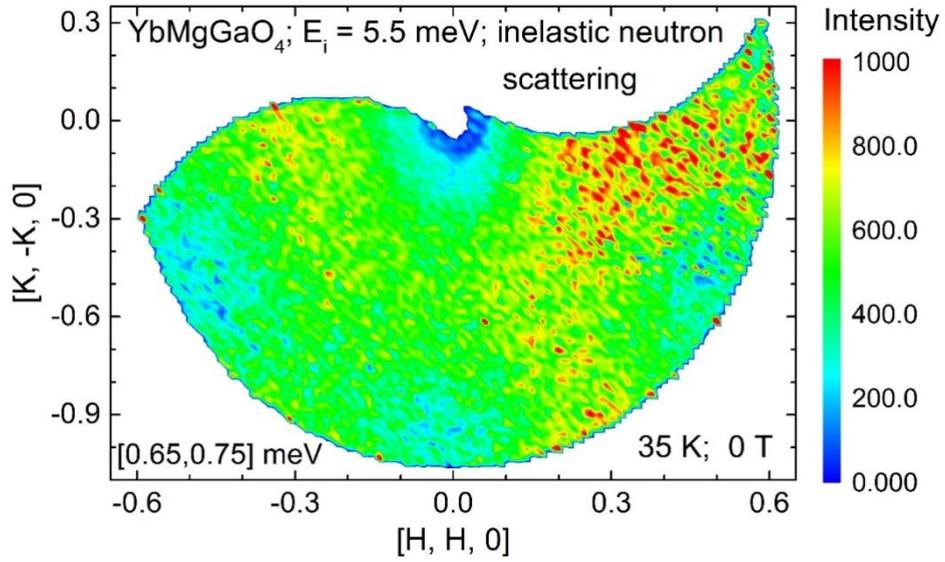

**Figure S7.** Raw inelastic neutron scattering for YbMgGaO$_4$ at 35 K and 0 T using the incident neutron energy 5.5 meV. The data have been integrated over the momentum space, $-0.9 \leq \eta \leq 0.9$ in $[0, 0, -\eta]$, and over a small energy range, $0.65 \leq E \leq 0.75$ meV.

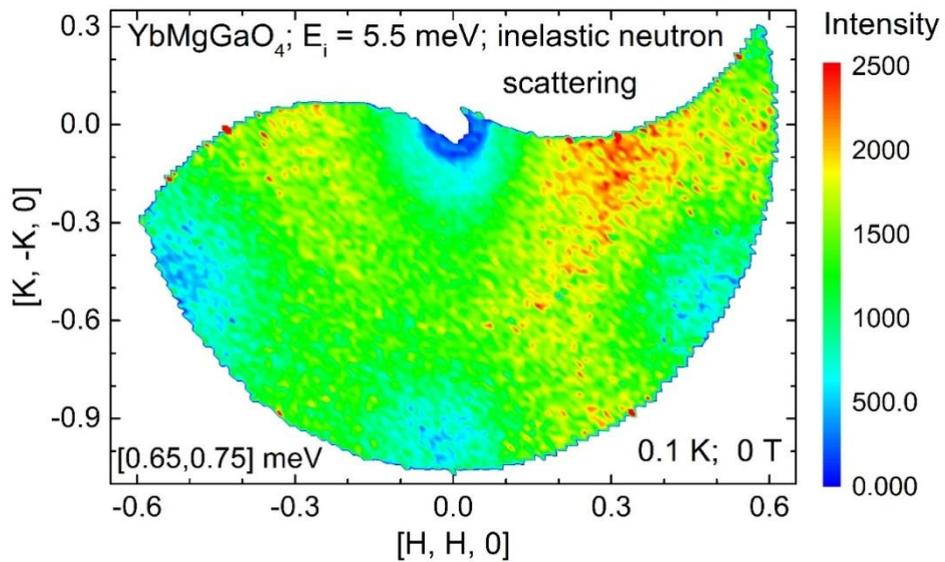

**Figure S8.** Raw inelastic neutron scattering for YbMgGaO$_4$ at 0.1 K and 0 T using the incident neutron energy 5.5 meV. The data have been integrated over the momentum space, $-0.9 \leq \eta \leq 0.9$ in $[0, 0, -\eta]$, and over a small energy range, $0.65 \leq E \leq 0.75$ meV.

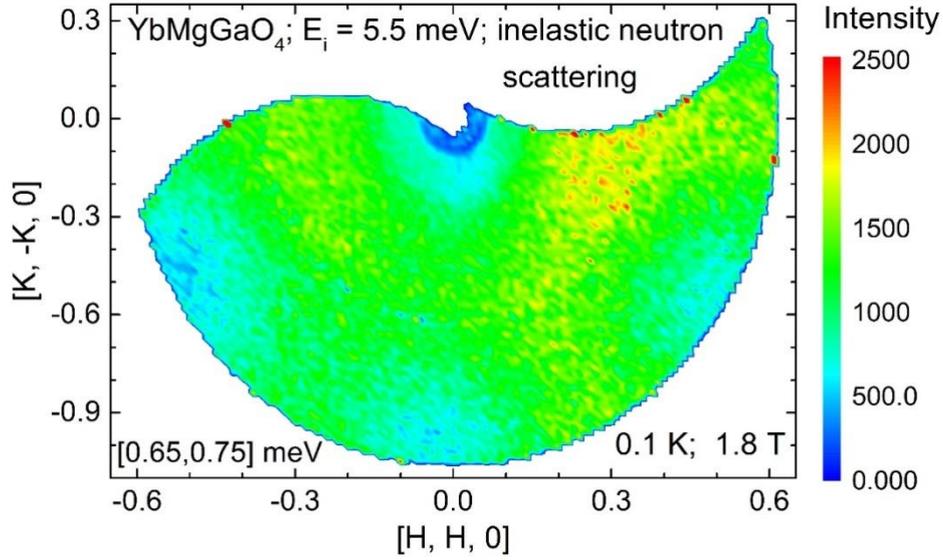

**Figure S9.** Raw inelastic neutron scattering for YbMgGaO$_4$ at 0.1 K and 1.8 T using the incident neutron energy 5.5 meV. The data have been integrated over the momentum space, $-0.9 \leq \eta \leq 0.9$ in $[0, 0, -\eta]$, and over a small energy range, $0.65 \leq E \leq 0.75$ meV.

In Fig. 1f of the main text, as well as in Fig. 1a and b, the coordinate point $(x, y)$ represents the wave-vector, $x[1, 1, 0] + y[1, -1, 0] = [x+y, x-y, 0] = (x+y)\boldsymbol{\beta_1}+(x-y)\boldsymbol{\beta_2}+0\boldsymbol{\beta_3}$ (here, $\boldsymbol{\beta_1}$, $\boldsymbol{\beta_2}$ and $\boldsymbol{\beta_3}$ are the inverted lattice vectors). That is, H = $x+y$ and K = $x-y$ in Eq. (2) (see the main text).

The similar representation is used in other recent publications, such as Nature 540, 559–562 (2016) (triangular lattice) and Nature 492, 406-410 (2012) (kagome lattice). The corresponding equal-time INS intensity of uncorrelated nearest-neighbor valence bonds on a *kagome* lattice is proportional to,

$$|F(\mathbf{Q})|^2 = \frac{2}{3}|f(\mathbf{Q})|^2 \{3 - \cos(\pi H) - \cos(\pi K) - \cos[\pi(H+K)]\}. \qquad \text{Eq. S2}$$

For herbertsmithite, $f(\mathbf{Q})$ stands for the magnetic form factor of free $Cu^{2+}$, Eq. S2 is exactly consistent with the calculations previously reported in the reference, Nature 492, 406-410 (2012).

The equal-time INS intensity for uncorrelated next-nearest-neighbor valence bonds on the triangular lattice is proportional to,

$$|F(\mathbf{Q})|^2 = \frac{2}{3}|f(\mathbf{Q})|^2 \{3 - \cos[2\pi(2H+K)] - \cos[2\pi(H+2K)] - \cos[2\pi(H-K)]\}. \qquad \text{Eq. S3}$$

The equal-time INS intensity for uncorrelated third-nearest-neighbor valence bonds on the triangular lattice is proportional to,

$$|F(\mathbf{Q})|^2 = \frac{2}{3}|f(\mathbf{Q})|^2 \{3 - \cos(4\pi H) - \cos(4\pi K) - \cos[4\pi(H+K)]\}. \qquad \text{Eq. S4}$$

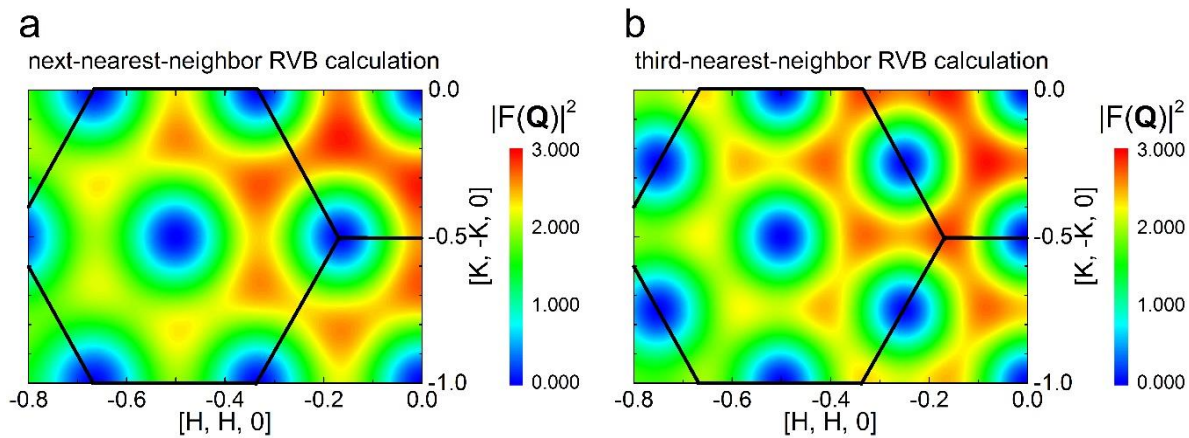

**Figure S10.** Calculated equal-time scattering intensity on the triangular lattice. **a**, calculated inelastic neutron scattering intensity for uncorrelated next-nearest-neighbor valence bonds (Eq. S3), and, **b**, for uncorrelated third-nearest-neighbor valence bonds (Eq. S4). The black lines represent Brillouin zone boundaries. Both calculations are clearly very different from our experimental observations (see Fig. S7 and Fig. S8).

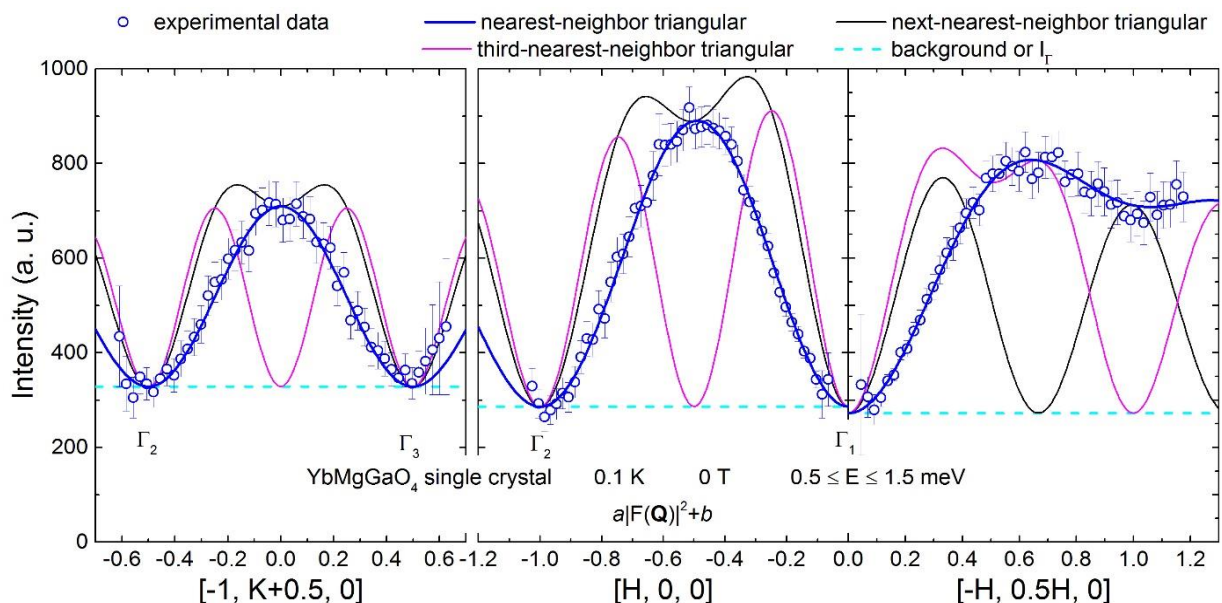

**Figure S11** Wave-vector dependence of the INS intensity along [-1, K+0.5, 0], [H, 0, 0], and [-H, 0.5H, 0], with lines representing different RVB calculations, $a|F(\mathbf{Q})|^2+b$. Here, $a$ and $b$ are fitting parameters, the pre-factor and background constant, respectively. The **Q**-dependence of the signal follows the nearest-neighbor RVB model on the triangular lattice very well, while it is inconsistent with other valence bond correlations. Please note: Due to the unequal number of integrated data points, both $a$ and $b$ are slightly different along different **Q** directions. And there are slight boundary effects at the beginning and end of each series of the experimental data.

**Table S1**. The fitted pre-factors *a* and background constants *b* at 35 and 0.1 K using the nearest-neighbor RVB model on the triangular lattice.

| Directions | [-1, K+0.5, 0] (Fig.1c) | | [H, 0, 0] (Fig.1d) | | [-H, 0.5H, 0] (Fig.1e) | |
|---|---|---|---|---|---|---|
| Temperature | 35 K | 0.1 K | 35 K | 0.1 K | 35 K | 0.1 K |
| (2/3)*a* | 29.1(1.7) | 109.6(2.7) | 50.9(2.2) | 158.3(4.6) | 42.5(2.1) | 126.4(3.9) |
| *a*(35K)/*a*(0.1K) | 0.266(22) | | 0.322(23) | | 0.336(27) | |
| *b* or $I_\Gamma$ | 165.4(3.2) | 328.6(4.9) | 130.7(4.3) | 286.1(8.5) | 118.6(5.1) | 272.5(9.3) |
| *b*(35K)/*b*(0.1K) or $I_\Gamma(35K)/I_\Gamma(0.1K)$ | 0.503(17) | | 0.457(29) | | 0.435(34) | |

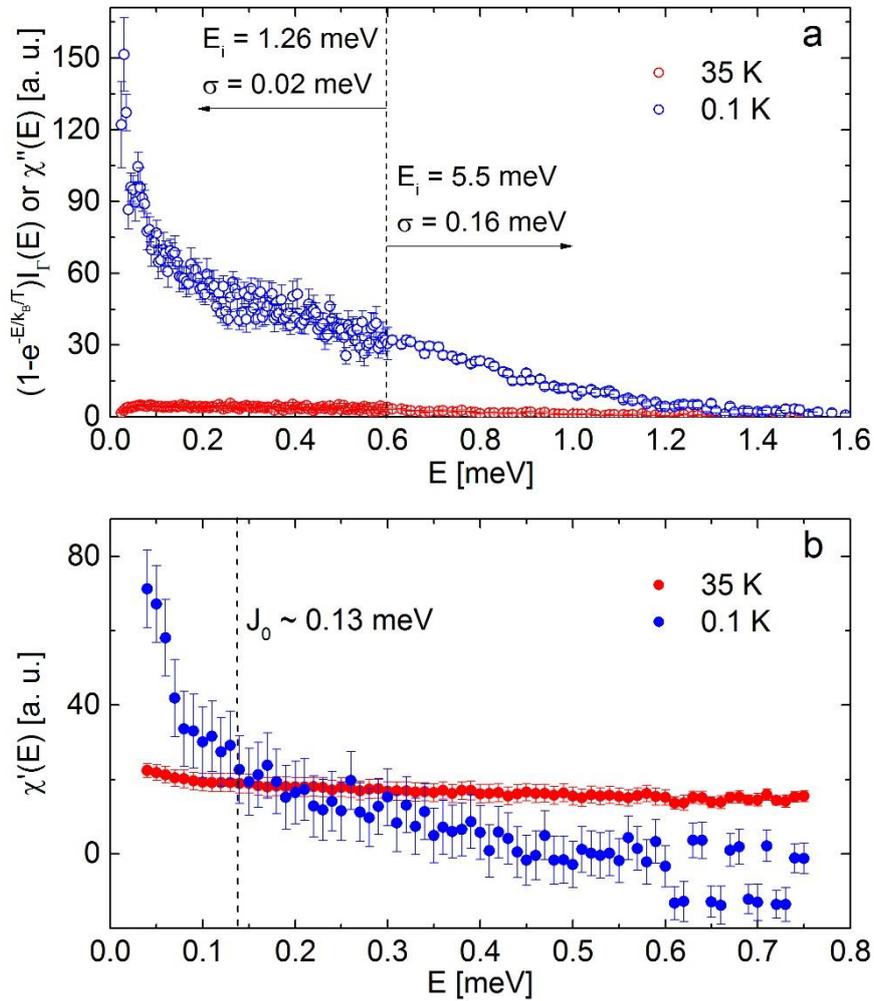

**Figure S12** Energy dependence of the uniform susceptibility (**Q** ~ 0) measured at 0.1 and 35 K. **a**, imaginary part of the susceptibility; **b** real part of the susceptibility. The elastic signal determined by fitting the spectrum at $|E| \leq \sigma$ was subtracted from $I_\Gamma(E)$. The spectra at 0.02 < E < 0.6 meV were obtained from the measurements with $E_i$ = 1.26 meV ($\sigma$ = 0.02 meV), while the spectra at E > 0.6 meV were obtained from the measurements with $E_i$ = 5.5 meV ($\sigma$ = 0.16 meV). The integration in the Kramers-Kronig relation was performed starting from 0.02 meV, and the data points with |E'-E| < 0.005 meV were excluded in the integral to avoid the divergence (see Eq. S6).

At the Gamma point ($\Gamma_1$: K = 0; H = 0, and **Q** = 0), the measured INS intensity, $I_\Gamma(E)$, is related to the uniform spin susceptibility, $\chi'(E)$. Via the fluctuation-dissipation theorem, we obtain imaginary part of the susceptibility [Rev. Sci. Instrum. 84, 083906 (2013)],

$$\chi''(E) \propto \left(1 - e^{-\frac{E}{k_B T}}\right) I_\Gamma(E), \qquad \text{Eq. S5}$$

assuming the Debye-Waller factor is temperature-independent below ~ 40 K.

Real part of the susceptibility can be obtained through the Kramers-Kronig transformation,

$$\chi'(E) \propto \frac{1}{\pi} \int \frac{\left(1 - e^{-\frac{E'}{k_B T}}\right) I_\Gamma(E')}{E' - E} dE'. \qquad \text{Eq. S6}$$

Energy dependence of the uniform (**Q** ~ 0) susceptibility is shown in Fig. S12. Experimental INS intensity ($I_\Gamma$) at 0.1 K is significantly larger than that at 35 K, $I_\Gamma$(0.1 K) ~ $2I_\Gamma$(35 K) (0.5 ≤ E ≤ 1.5 meV). Nevertheless, above 0.5 meV real part of the susceptibility ($\chi'$) measured at 0.1 K turns out to be zero within the error bar, which is well in line with our RVB scenario. High values of $\chi'$ are seen at energies below ~ 0.15 meV only, indicating some low-energy physics beyond the simple RVB picture.

The comparison of $\chi'$ with the bulk susceptibility (Fig. S13) indicates that the susceptibility at 0.48 K is higher than the one expected at 0.1 K, even at low energies, where $\chi'(E)$ is finite. Therefore, we anticipate a downward trend in the susceptibility around 0.1 K and exclude any ferromagnetic instability in the system.

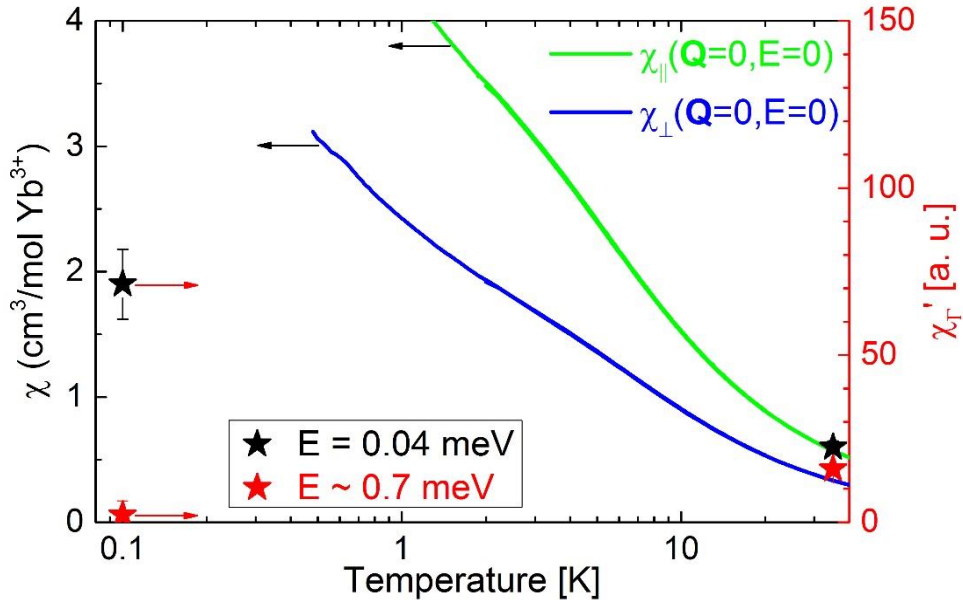

**Figure S13** Temperature dependences of the static bulk susceptibilities measured with the field parallel ($\chi_\parallel$) and perpendicular ($\chi_\perp$) to the *c*-axis for the YbMgGaO$_4$ single crystal [Phys. Rev. Lett. 117, 097201 (2016)]. The black stars show the uniform spin susceptibilities at E = 0.04 meV, while the red stars show the uniform spin susceptibilities around E = 0.7 meV, as extracted from the INS data.

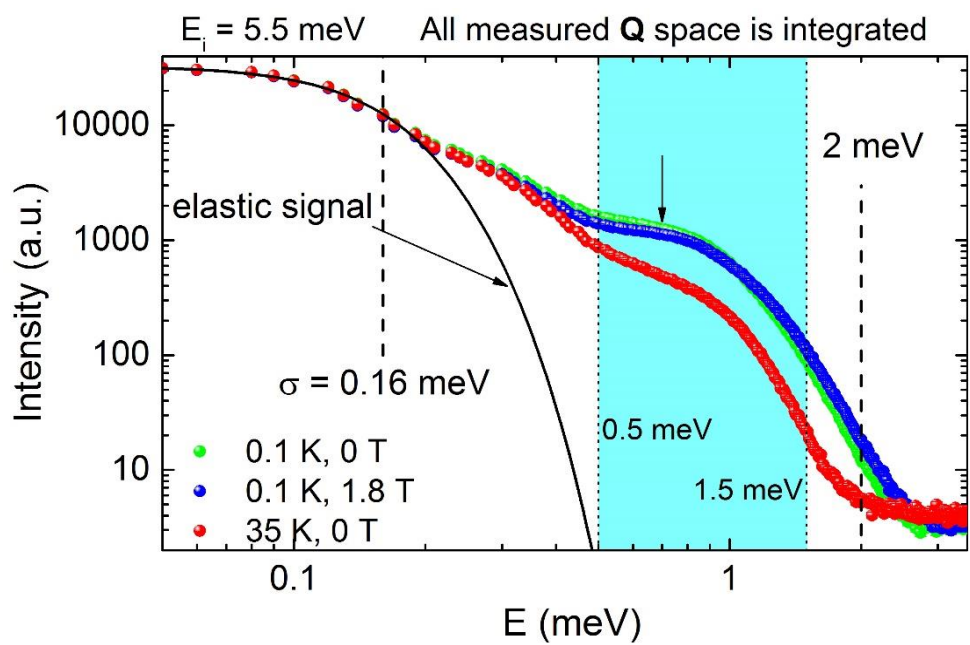

**Figure S14** Energy dependence of the integrated neutron scattering intensity for YbMgGaO$_4$ with E$_i$ = 5.5 meV. Solid line represents the elastic contribution.

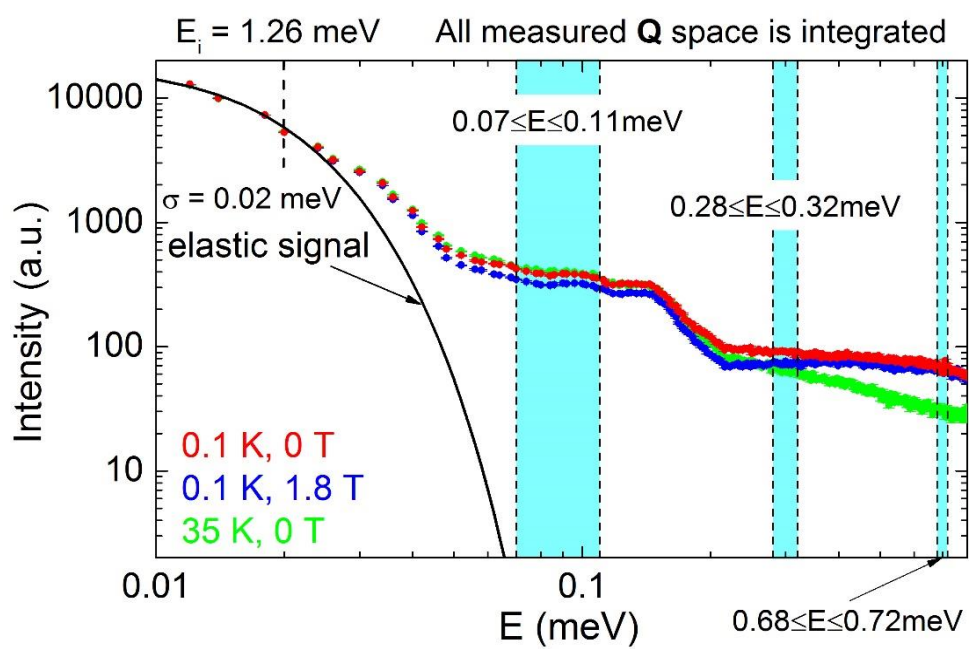

**Figure S15** Energy dependence of the integrated neutron scattering intensity, and the elastic component (E$_i$ = 1.26 meV).

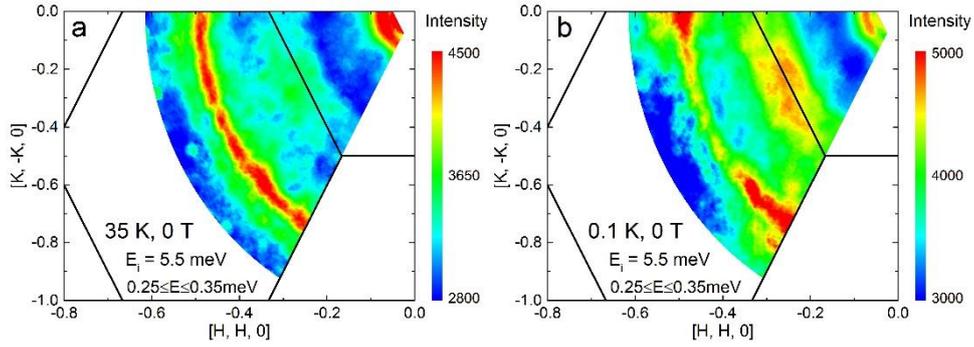

**Figure S16** Wave-vector dependence of the INS intensity ($0.25 \leq E \leq 0.35$ meV, with $E_i = 5.5$ meV) measured **a**, at 35 K and 0 T, **b**, at 0.1 K and 0 T. They are contaminated by a polycrystalline-like signal of some sort. It may originate from the magnet, cryostat, and/or sample holder.

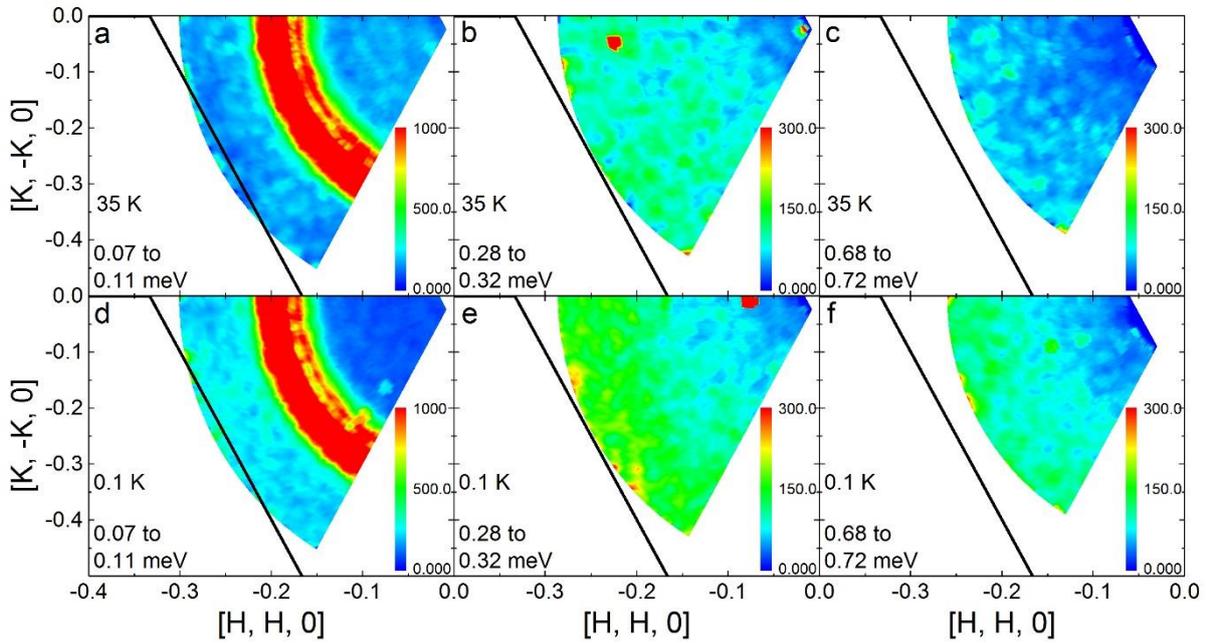

**Figure S17** Wave-vector dependence of the INS intensity measured under 0 T with $E_i = 1.26$ meV **a**, at $0.07 \leq E \leq 0.11$ meV and 35 K, **b**, at $0.28 \leq E \leq 0.32$ meV and 35 K, **c**, at $0.68 \leq E \leq 0.72$ meV and 35 K, **d**, at $0.07 \leq E \leq 0.11$ meV and 0.1 K, **e**, at $0.28 \leq E \leq 0.32$ meV and 0.1 K, and **f**, at $0.68 \leq E \leq 0.72$ meV and 0.1 K. There are no qualitative differences between the spectra at ~ 0.3 and ~ 0.7 meV in all measured **Q** space apart from an increase in the intensity (please compare b with c at 35 K and e with f at 0.1 K).

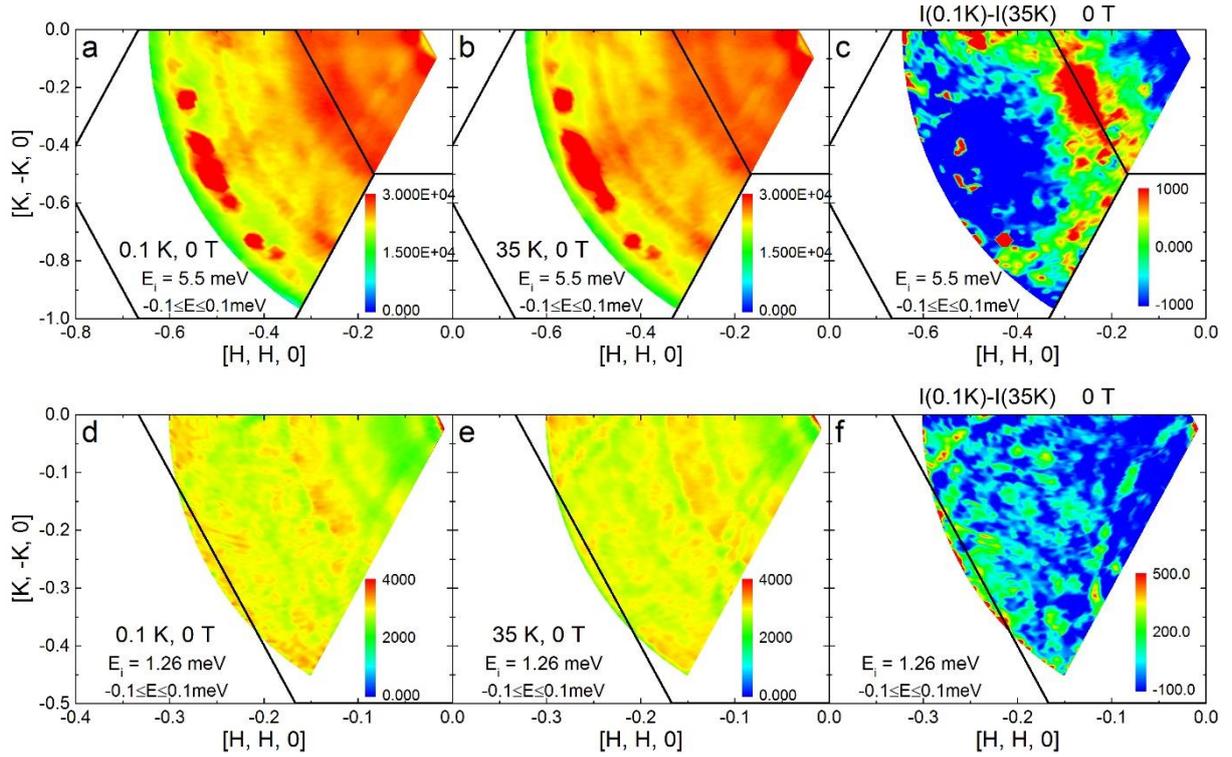

**Figure S18** Wave-vector dependence of the elastic scattering signal (integrated over $-0.1 \leq E \leq 0.1$ meV) **a**. with $E_i = 5.5$ meV at 0.1 K, **b**. with $E_i = 5.5$ meV at 35 K, **d**. with $E_i = 1.26$ meV at 0.1 K, and **e**. with $E_i = 1.26$ meV at 35 K. I(0.1 K) - I(35 K) **c**. with $E_i = 5.5$ meV, and **f**. with $E_i = 1.26$ meV. The polycrystalline signal of some sort seriously contaminates the intrinsic signal of YbMgGaO$_4$. We can subtract the 0.1 K signal with the 35 K signal, but the subtracted signals become significantly negative and seem too noisy in some regions of the **Q** space (see c and f), and the measured **Q** space is also very limited with $E_i = 1.26$ meV (see f).

The positive signal in I(0.1K) - I(35K) at $|E| \leq 0.1$ meV around the M-point is likely intrinsic (see Fig. S18 c) and seems largely consistent with the earlier INS data by Paddison *et al.* [Nat. Phys. 13, 117-122 (2017)] who observed maxima of the diffuse scattering at the M-points, as opposed to the maxima at the K-points above 0.5 meV. This shift of the intensity reflects a fundamental change from the predominant nearest-neighbor correlations at high energies to long-range correlations that manifest themselves at very low energies only. Detailed analysis of these long-range correlations is hindered by the contamination of our low-energy data with elastic signals from the sample environment.

Whereas the full energy range starting from 0 meV is unfeasible for the analysis, we argue that the **Q**-dependence of the intensity obtained by integrating the data above 0.5 meV is sufficient for drawing robust conclusions on the nature of spin-spin correlations underlying the excitation continuum. First, our analysis is performed along the lines of the recent INS study of herbertsmithite, where lower-energy spectral weight was also excluded. Specifically, for herbertsmithite the spectra were integrated starting from $E_{min} = 1$ meV, which is *well above* the energy resolutions (1/2σ), 0.21 ($E_{min}/\sigma = 2.4$) and 0.08 meV ($E_{min}/\sigma = 6.3$) [half-width at half-maximum, please see METHODS SUMMARY in Nature 492, 406-410 (2012)]. In our case, we use the same strategy and adopt $E_{min}/\sigma = 0.5/0.16 = 3.1$ to completely exclude the elastic signal (see Fig. S14).

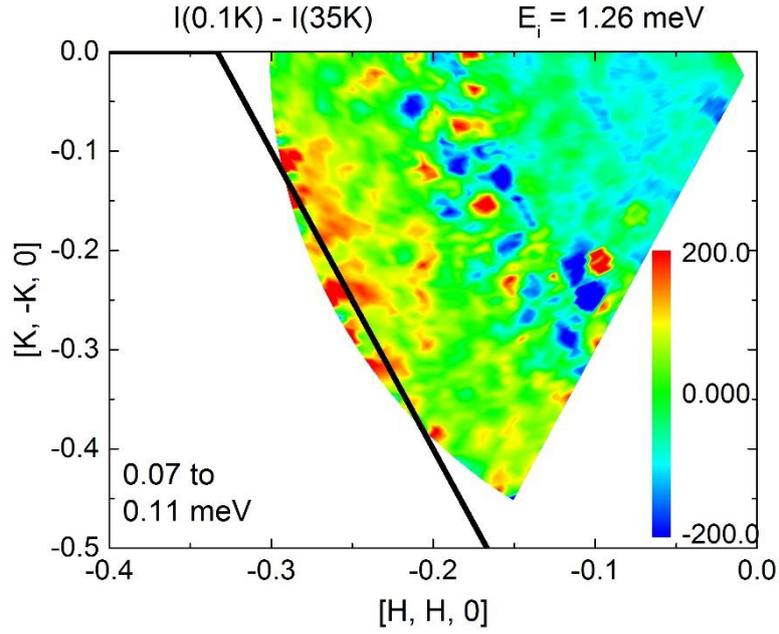

**Figure S19** Wave-vector dependences of the subtracted INS intensity [I(0.1K) – I(35K)] integrated over $0.07 \leq E \leq 0.11$ meV with $E_i = 1.26$ meV. The subtracted signals become significantly negative in some regions of the **Q** space.

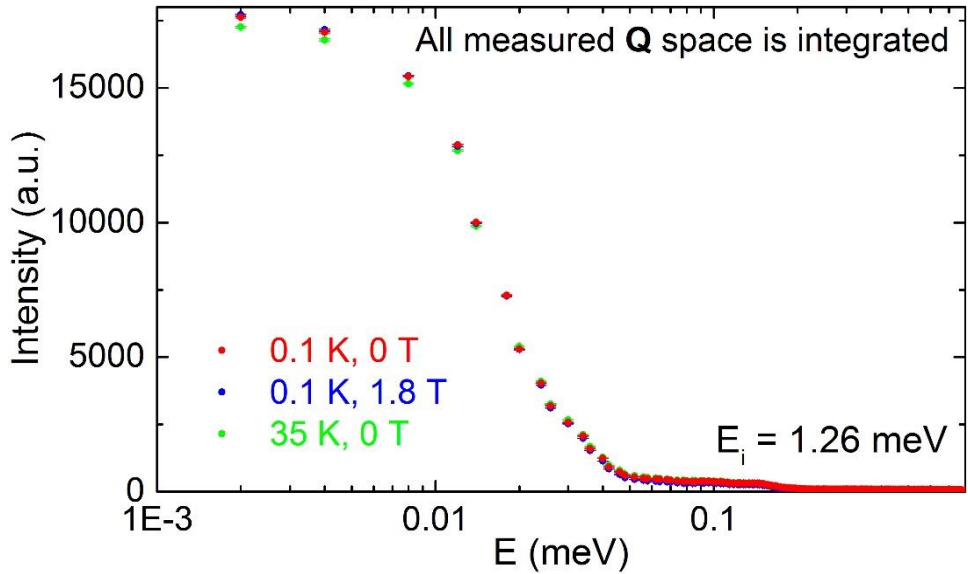

**Figure S20.** Raw INS data for YbMgGaO$_4$ with the lowest incident neutron energy of 1.26 meV. The data have been integrated over all measured **Q** space.

Further on, we verified that the spectral weight about 0.5 meV bears characteristic signatures of the ground state. To this end, we considered an alternative state, the 120-deg long-range order, and calculated spin-wave excitations using experimental parametrization of the spin Hamiltonian of YbMgGaO$_4$ ($J_{zz} = 0.085$ meV, $J_{\pm} = 0.078$ meV, $J_{\pm\pm} = 0.013$ meV and $J_{z\pm} = 0.003$ meV) [Phys. Rev. Lett. 115, 167203 (2015)]. Magnetic excitations calculated on the level of the linear spin-wave theory [J. Phys. Condens. Matter 27, 166002 (2015)] (see Fig. S21), were broadened with the energy resolution of the LET spectrometer, $\sigma = 0.16$ meV (see Fig. S22).

We integrate the calculated spectrum from half the spin-wave bandwidth (~ 0.2 meV) to energies well above the top of the band (~ 0.5 meV) (see Fig. S22). The magenta curves in Fig. S23 show that the profile of the integrated spin-wave spectrum is clearly inconsistent with the experimental one. For the sake of completeness, we also show the spin-wave spectra integrated from 0, 0.12, 0.3, 0.34 and 0.4 meV. None of them is qualitatively consistent with the experimental data (see Fig. S23).

A similar result can be obtained for the nearest-neighbor Heisenberg Hamiltonian on the triangular lattice that also yields the 120-deg long-range-ordered state. We thus conclude that even at high energies excitations from the 120-deg ordered state are distinct from those of the RVB state. Therefore, even above 0.5 meV (about 1/3 of the experimental bandwidth) the Q-dependence of the scattered intensity remains characteristic, and indicates predominant nearest-neighbor correlations, which are likely of the RVB type (see detailed discussion with points I–VI in the main text).

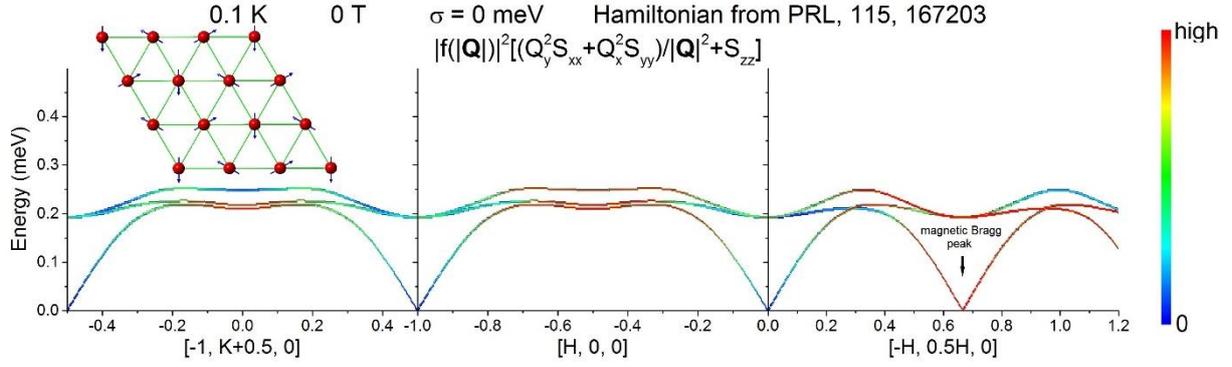

**Figure S21.** Calculated spin-wave excitations [Rev. Sci. Instrum. 84, 083906 (2013)] along [-1, K+0.5, 0], [H, 0, 0], and [-H, 0.5H, 0] for the hypothetical 120 degree magnetically ordered structure (see the inset for the magnetic unit cell) based on the experimental parametrization of the spin Hamiltonian of YbMgGaO$_4$ [Phys. Rev. Lett. 115, 167203 (2015)]. Here, f(|**Q**|) is the magnetic form factor of Yb$^{3+}$.

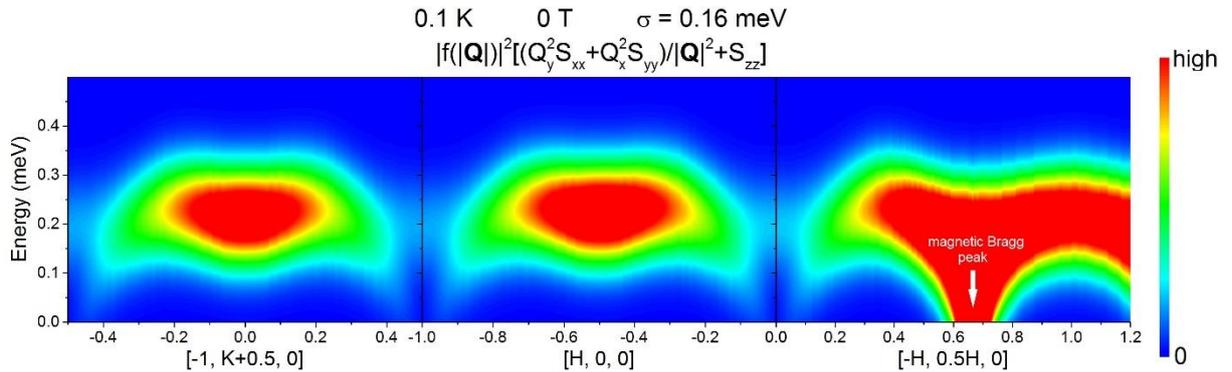

**Figure S22.** Calculated spin-wave excitations based on the experimental parametrization of the spin Hamiltonian (Fig. S21) with the Gaussian broadening, FWHM = 0.16 meV.

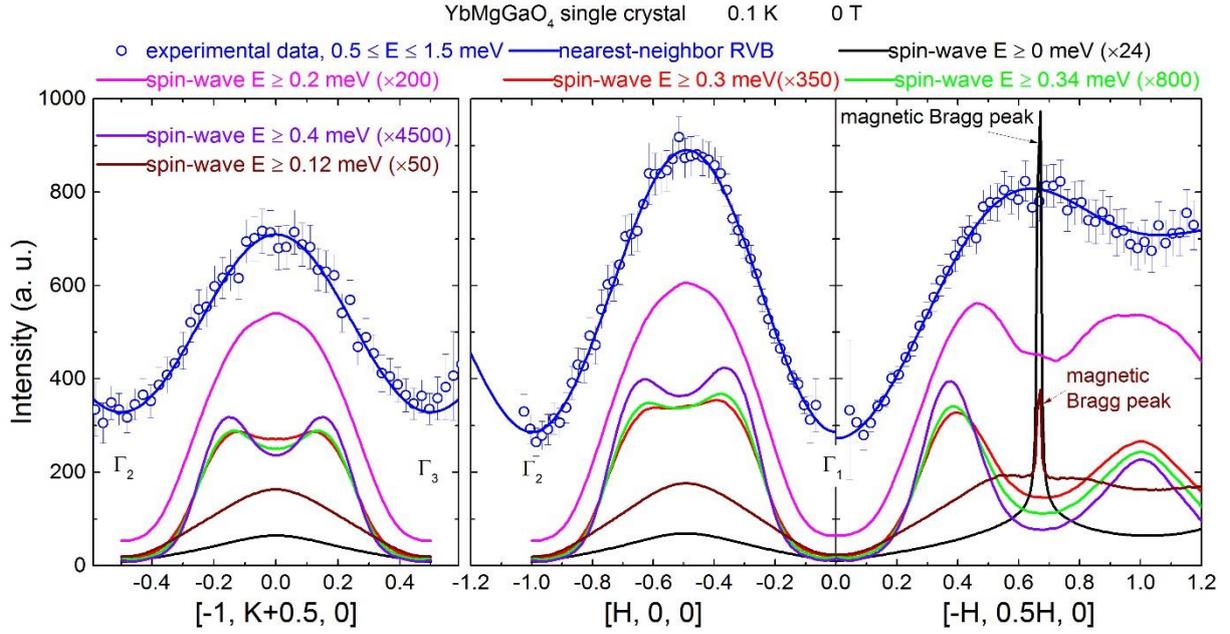

**Figure S23.** Wave-vector dependence of the INS intensity along [-1, K+0.5, 0], [H, 0, 0], and [-H, 0.5H, 0], with blue lines representing nearest-neighbor RVB calculations, $a|F(Q)|^2+b$. The spin-wave spectra based on the experimental parametrization of the spin Hamiltonian are integrated from 0, 0.12, 0.2, 0.3, 0.34, and 0.4 meV (all up to 0.5 meV) yielding the black, wine, magenta, red, green and violet curves, respectively. Above 0.3 meV, only minor changes in the **Q**-dependence are observed showing that even at very high energies the **Q**-dependence of the RVB state is not reproduced. The magnetic Bragg peak (-2/3, 1/3, 0), which is naturally present in the spin-wave spectra, reflects long-range order in the 120-deg state and disappears at high energies.

The calculated spin-wave excitations based on the Heisenberg model without any broadening are shown in Fig. S24. With an arbitrary Gaussian broadening, FWHM ~ 2J, the magnetic Bragg reflection (-2/3, 1/3, 0) of the periodic magnetic structure based on the Heisenberg model is very clear and is much stronger than that based on the YbMgGaO$_4$ Hamiltonian (see Fig. S25). To exclude this sharp Bragg peak reflecting the long-range correlations, we obtain the high-energy spin-wave excitations by integrating from 3.4J to 4J (see Fig. S26).

Both the calculated high-energy spin-wave spectra for the same magnetic structure indeed share almost the same profile, though the corresponding spin Hamiltonians are completely symmetrically different (see the pink and violet curves in Fig. S26). Both calculated high-energy spectra without magnetic Bragg reflections only reflect the short distance spin correlations of the 120 degree Neel state, while are essentially inconsistent with our measured high-energy INS spectrum (see Fig. S26).

The **Q**-dependence of the measured high-energy INS spectrum for YbMgGaO$_4$ cannot be understood by whatever short distance antiferromagnetic spin correlations on a triangular lattice, and the nearest-neighbor RVB scenario is still the most accurate explanation so far.

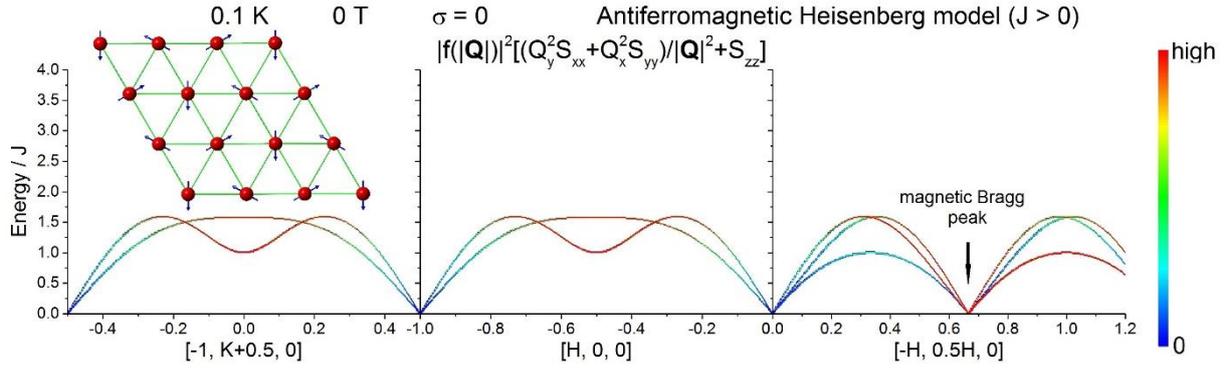

**Figure S24.** Calculated spin-wave excitations along [-1, K+0.5, 0], [H, 0, 0], and [-H, 0.5H, 0] for the 120 degree magnetically ordered ground state (see the inset for the magnetic unit cell) based on the nearest-neighbor spin-1/2 antiferromagnetic Heisenberg Hamiltonian on the triangular lattice (J > 0).

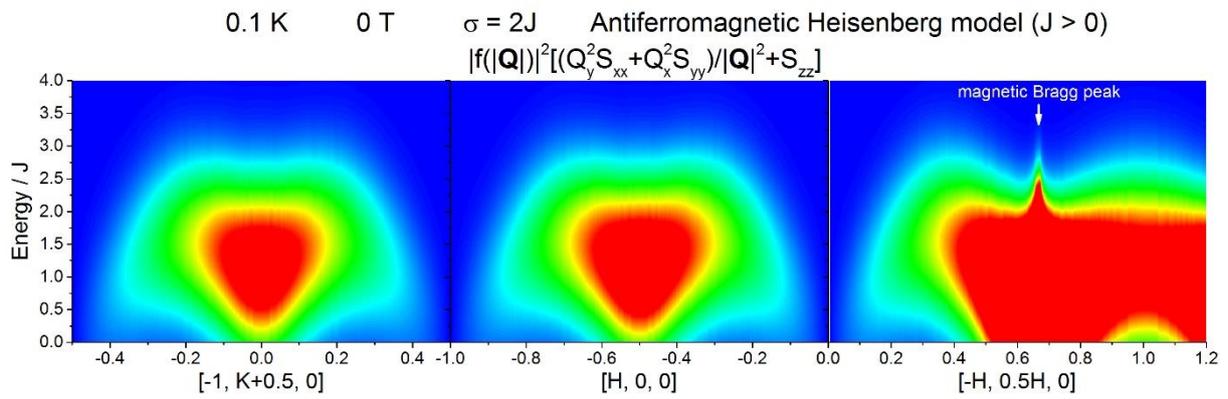

**Figure S25.** Calculated spin-wave excitations (Fig. S24) with the Gaussian broadening, FWHM = 2J.

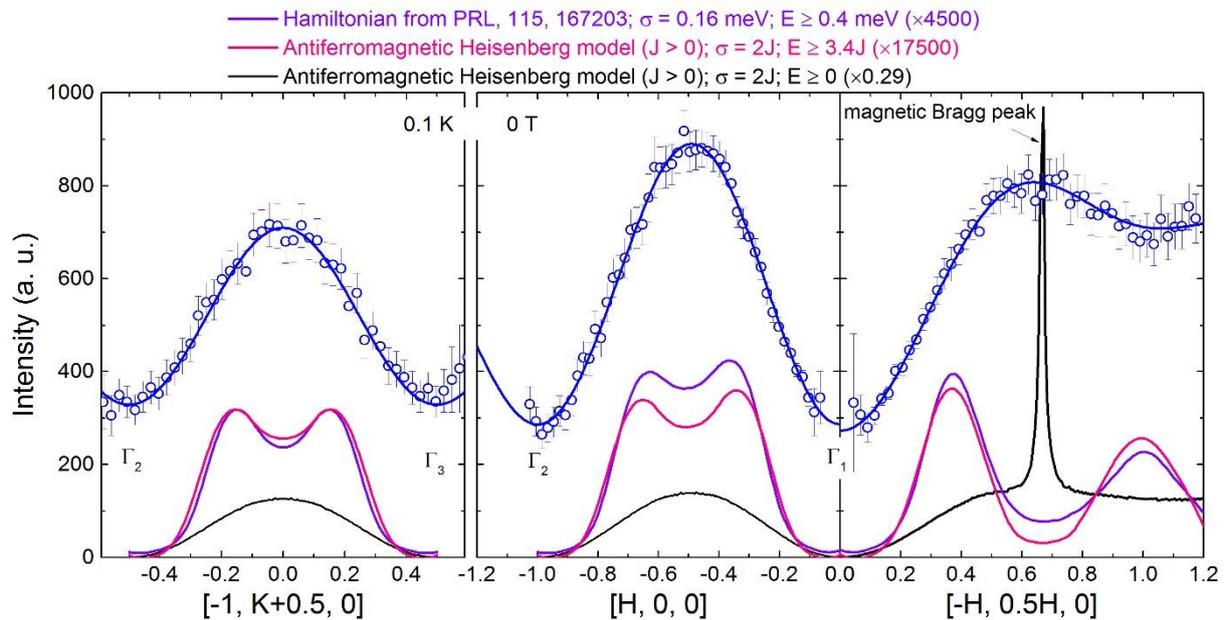

**Figure S26.** Wave-vector dependence of the INS intensity along [-1, K+0.5, 0], [H, 0, 0], and [-H, 0.5H, 0], with blue lines representing nearest-neighbor RVB calculations, $a|F(Q)|^2+b$. The spin-wave spectra for the nearest-neighbor Heisenberg Hamiltonian are integrated from 0J and 3.4J (both up to 4J) yielding the black and pink curves, respectively. The violet curves show the integration result for spin-wave excitations calculated for the YbMgGaO4 Hamiltonian (the violet curves in Fig. S23).